\documentclass[prd,12pt,superscriptaddress,preprintnumbers]{revtex4}
\usepackage{epsfig}
\usepackage{psfig}
\usepackage{graphicx}
\newcommand{\AHEP}{Instituto de F\'{\i}sica Corpuscular --
  C.S.I.C./Universitat de Val{\`e}ncia \\
  Edificio Institutos de Paterna, Apt 22085,
  E--46071 Val{\`e}ncia, Spain\\}
\begin{document} 

\title{Probing neutrino mass with multilepton production at the
  Tevatron in the simplest $R$-parity violation model}

\author{M.\ B.\ Magro}
\email{magro@fma.if.usp.br}
\affiliation{Instituto de F\'{\i}sica, 
             Universidade de S\~ao Paulo, S\~ao Paulo -- SP, Brazil.}
           
\author{F.\ de Campos}
\email{fernando@ift.unesp.br}
\affiliation{Departamento de F\'{\i}sica e Qu\'{\i}mica,
             Universidade Estadual Paulista - Brazil }

\author{O.\ J.\ P.\ \'Eboli}
\email{eboli@fma.if.usp.br}
\affiliation{Instituto de F\'{\i}sica, 
             Universidade de S\~ao Paulo, S\~ao Paulo -- SP, Brazil.}

\author{W.\ Porod} 
\email{porod@physik.unizh.ch} 
\affiliation{ Institut f\"ur Theoretische Physik, Universit\"at Z\"urich,
  Z\"urich, Switzerland }

\author{D.\ Restrepo}
\email{restrepo@uv.es}
\affiliation{Instituto de F\'{\i}sica,
              Universidad de Antioquia - Colombia}

\author{J.\ W.\ F.\ Valle} 
\email{valle@ific.uv.es}
\affiliation{\AHEP}

\begin{abstract}
  
  We analyse the production of multileptons in the simplest supergravity model
  with bilinear violation of $R$ parity at the Fermilab Tevatron. Despite the
  small $R$-parity violating couplings needed to generate the neutrino masses
  indicated by current atmospheric neutrino data, the lightest supersymmetric
  particle is unstable and can decay inside the detector. This leads to a
  phenomenology quite distinct from that of the $R$-parity conserving
  scenario.  We quantify by how much the supersymmetric multilepton signals
  differ from the $R$-parity conserving expectations, displaying our results
  in the $m_0 \otimes m_{1/2}$ plane. We show that the presence of bilinear
  $R$-parity violating interactions enhances the supersymmetric multilepton
  signals over most of the parameter space, specially at moderate and large
  $m_0$.

\vskip 24pt

%\textbf{DRAFT 08; 15/04/2003 }

\end{abstract}

\preprint{ ZU-TH 04/03 }

\maketitle
\newpage

%%%%%%%%%%%%%%%%%%%%%%%%%%%%%%%%%%%%%%%%%%%%%%%%%%%%%%%%%%%%%%%%%%%%%%
\section{Introduction}

Supersymmetry (SUSY) provides a promising candidate for physics beyond the
Standard Model (SM). The search for supersymmetric partners of the SM
particles constitutes an important item in the agenda of current high energy
colliders like the Tevatron, and future colliders like the CERN Large Hadron
Collider or a linear $e^+ e^-$ collider.  So far most of the effort in
searching for supersymmetric signatures has been confined to the framework of
$R$-parity conserving realizations~\cite{susywg}; see, however,
Ref.~\cite{Allanach:1999bf} and references therein.

Recent data on solar and atmospheric neutrinos give a robust evidence for
neutrino conversions~\cite{Fukuda:2002pe}, probably the most profound
discovery in particle physics in the recent years.  It has been suggested long
ago that neutrino masses and supersymmetry may be deeply tied
together~\cite{rpold}.  Indeed, SUSY models exhibiting $R$-parity violation
can lead to neutrino masses and mixings~\cite{Hirsch:2000ef} in agreement with
the current solar and atmospheric neutrino data.  Furthermore, the simplest
possibility is bilinear $R$-parity violation ~\cite{bilinear,Diaz:1997xc}
which may arise as an effective description of a spontaneous $R$-parity
violation scenario~\cite{Masiero:1990uj}, or from some suitable \texttt{ab
  initio} symmetries~\cite{Mira:2000gg}.

It is interesting to notice that neutrino mass models based on $R$-parity
violation can be tested at colliders~
\cite{Bartl:2000yh,Porod:2000hv,Hirsch:2002ys}.  In this work, we study the
production of multileptons ($\ge 3 \ell$ with $\ell = e$ or $\mu$) at the
Fermilab Tevatron within the framework of the simplest supergravity (SUGRA)
model without $R$-parity conservation~\cite{Diaz:1997xc}.  We show that the
presence of bilinear $R$-parity violating (BRpV) interactions enhances the
supersymmetric multilepton signals over most of the parameter space, specially
at moderate and large $m_0$. In order to make the comparison with the
$R$-parity conserving case easier, we performed our detailed study of the
signal and SM backgrounds, adopting the same cuts (soft cuts SC2) proposed in
Ref.\ \cite{tata1}.

In SUGRA with universal soft breaking terms at unification, the masses of the
sleptons, the lighter chargino ($\tilde{\chi}^\pm_1$), and the lighter
neutralinos ($\tilde{\chi}^0_1$ and $\tilde{\chi}^0_2$) are considerably
smaller than the gluino and squark ones over a large range of the parameter
space \cite{tata1}.  Therefore, the production of charginos and neutralinos
provides the largest reach at the Tevatron.  In $R$-parity conserving
scenarios, a promising signal for SUGRA at the Tevatron is the production of
$\tilde{\chi}^0_2 \tilde{\chi}^\pm_1$ pairs and their subsequent decay into
three charged leptons in association with missing energy due to the undetected
lightest supersymmetric particle (LSP), which turns out to be
$\tilde{\chi}^0_1$.  In the presence of $R$-parity violation, the LSP is no
longer stable, giving rise to events containing more particles, and it can be
electrically charged since it decays. Interesting signals for such scenarios
have been worked out for staus~\cite{Hirsch:2002ys},
stops~\cite{Bartl:1996gz}, gluinos~\cite{Bartl:1996cg} and also for
supersymmtric Higgs bosons~\cite{Akeroyd:1997iq}.  The larger activity in the
detector can either enhance the signal, via the production of additional
leptons or suppress it due to the hadronic activity that spoils the isolation
of the leptons. Therefore, a careful reanalysis of the trilepton signal is
necessary.

In this work, we consider a SUGRA model that includes the following bilinear
terms in the superpotential~\cite{bilinear,Diaz:1997xc,bilinear1}
\begin{equation}
W_{\text{BRpV}} = W_{\text{MSSM}}  - \varepsilon_{ab}
\epsilon_i \widehat L_i^a\widehat H_u^b \; ,
\end{equation}
where the last term violates $R$ parity. In order to reproduce the values of
neutrino masses indicated by current neutrino data~\cite{Maltoni:2002ni} we
must have $|\epsilon_i|\ll|\mu|$, where $\mu$ denotes the SUSY bilinear mass
parameter~\cite{Hirsch:2000ef}.  The relevant bilinear terms in the soft
supersymmetry breaking sector are
\begin{equation}
V_{\text{soft}} = m_{H_u}^2H_u^{a*}H_u^a+m_{H_d}^2H_d^{a*}H_d^a+
M_{L_i}^2\widetilde L_i^{a*}\widetilde L_i^a -\varepsilon_{ab}\left(
B\mu H_d^aH_u^b+B_i\epsilon_i\widetilde L_i^aH_u^b\right) \; ,
\end{equation}
where the terms proportional to $B_i$ are the ones that violate $R$ parity.
The explicit $R$-parity violating terms induce vacuum expectation values (vev)
$v_i$, $i=1,2,3$ for the sneutrinos, in addition to the two Higgs vev's $v_u$
and $v_d$.

Our goal is to determine the impact of $R$-parity violation on SUSY
multilepton signals, for example in the production of three or more electrons
or muons, taking into account the magnitude of the mass scale indicated by the
current atmospheric neutrino experiments. In phenomenological studies where
the details of the neutrino sector are not relevant, it has been proven very
useful to work in the approximation where $R$ parity and lepton number are
violated in only one generation \cite{Akeroyd:1997iq,BRpB_tau}.  Thus, the
first step to achieve our goal is to assume the approximation where $R$ parity
is violated only in the third generation (of course all gauge interactions are
treated in the full three-generation scheme).  This is the theoretically
natural choice to make, since the third generation forms the basis for the
radiative breaking of the electroweak symmetry, driven by the top quark Yukawa
coupling. From the point of view of the analysis presented below, the breaking
of $R$ parity only in the third generation also corresponds to the
worst-case-scenario: the breaking of $R$ parity in the muon channel would
produce muons directly, not just as tau decay products, leading to an enhanced
multilepton (multi-muon) signal.

%%%%%%%%%%%%%%%%%%%%%%%%%%%%%%%%%%%%%%%%%%%%%%%%%%%%%%%%%%%%%%%%%%%%%%
\section{ Main features of our BRpV model}

The parameter space of our SUGRA model, which exhibits $R$-parity violation
only in the third generation, is
\begin{equation}
m_0\,,\, m_{1/2}\,,\, \tan\beta\,,\, {\mathrm{sign}}(\mu)\,,\, 
A_0 \,,\, 
\epsilon_\tau \: {\mathrm{, and}}\,\, m_{\nu_3}\,,
\end{equation}
where $m_{1/2}$ and $m_0$ characterise the common gaugino mass and scalar soft
SUSY breaking masses at the unification scale, $A_0$ is the common trilinear
term, and $\tan\beta$ is the ratio between the Higgs field vev's.  Although
many parameterisations are possible, it is convenient to characterise the BRpV
sector by the bilinear superpotential term $\epsilon_\tau$ and the neutrino
mass $m_{\nu_3}$.

%%%aqui

We considered the running of the masses and couplings to the electroweak
scale, assumed to be the top mass, using the one--loop renormalization group
equations. In the evaluation of the gaugino masses, we included the
next--to--leading order (NLO) corrections coming from $\alpha_s$, the
two--loop top Yukawa contributions to the beta-functions, and threshold
corrections enhanced by large logarithms; for details see \cite{GGW}. The NLO
corrections are especially important for $M_2$, leading to a change in the
wino mass up to 20\%. We then input the soft terms into PYTHIA that was used
to evaluated the masses and decay rates of all particles, except the first and
second neutralinos.

%%%

Present neutrino oscillation data fix the mass splittings among the three
neutrinos, leaving arbitrary the overall scale of neutrino masses. The latter
could be as large as an electron volt or so without conflicting with
cosmology, beta decays and neutrinoless double beta decays, given current
uncertainties in nuclear matrix element calculations~\cite{Valle:2002tm}.
However, in the BRpV model, neutrino masses are strongly
hierarchical~\cite{Hirsch:2000ef}, specially the lightest neutrino mass. As a
result the possibility of quasi-degenerate neutrinos is not realized in this
model and, correspondingly, we will not discuss it any further.  In what
follows, whenever we fix the BRpV parameters, we take the tree level neutrino
mass as $m_{\nu_3}=0.05$ eV, the current atmospheric best fit value from
Ref.~\cite{Maltoni:2002ni,Gonzalez-Garcia:2002dz}, and fix the value of the
remaining BRpV parameter $\epsilon_\tau$ at representative values 0.22 GeV and
$7 \times 10^{-4}$ GeV.

The presence of BRpV induces a mixing between the neutrinos and neutralinos,
giving rise to $R$-parity violating decays of the LSP. In our model, the
lightest neutralino presents leptonic decays $ \tilde{\chi}^0_1 \to \nu \ell^+
\ell^{\prime-}$, semi-leptonic ones $ \tilde{\chi}^0_1 \to \nu q \bar q $ or $
\ell q \bar q$, and the invisible mode $\tilde{\chi}^0_1 \to \nu \nu \nu$
\cite{Porod:2000hv}.  The expected $\tilde{\chi}^0_1$ lifetime and decay
lengths depend both on the magnitude of $R$-parity breaking parameters and the
chosen values of the SUGRA parameters.

\begin{figure*}[thbp] 
\includegraphics[height=9cm,width=0.46\linewidth,clip]{dec1.eps}
\includegraphics[height=9cm,width=0.46\linewidth,clip]{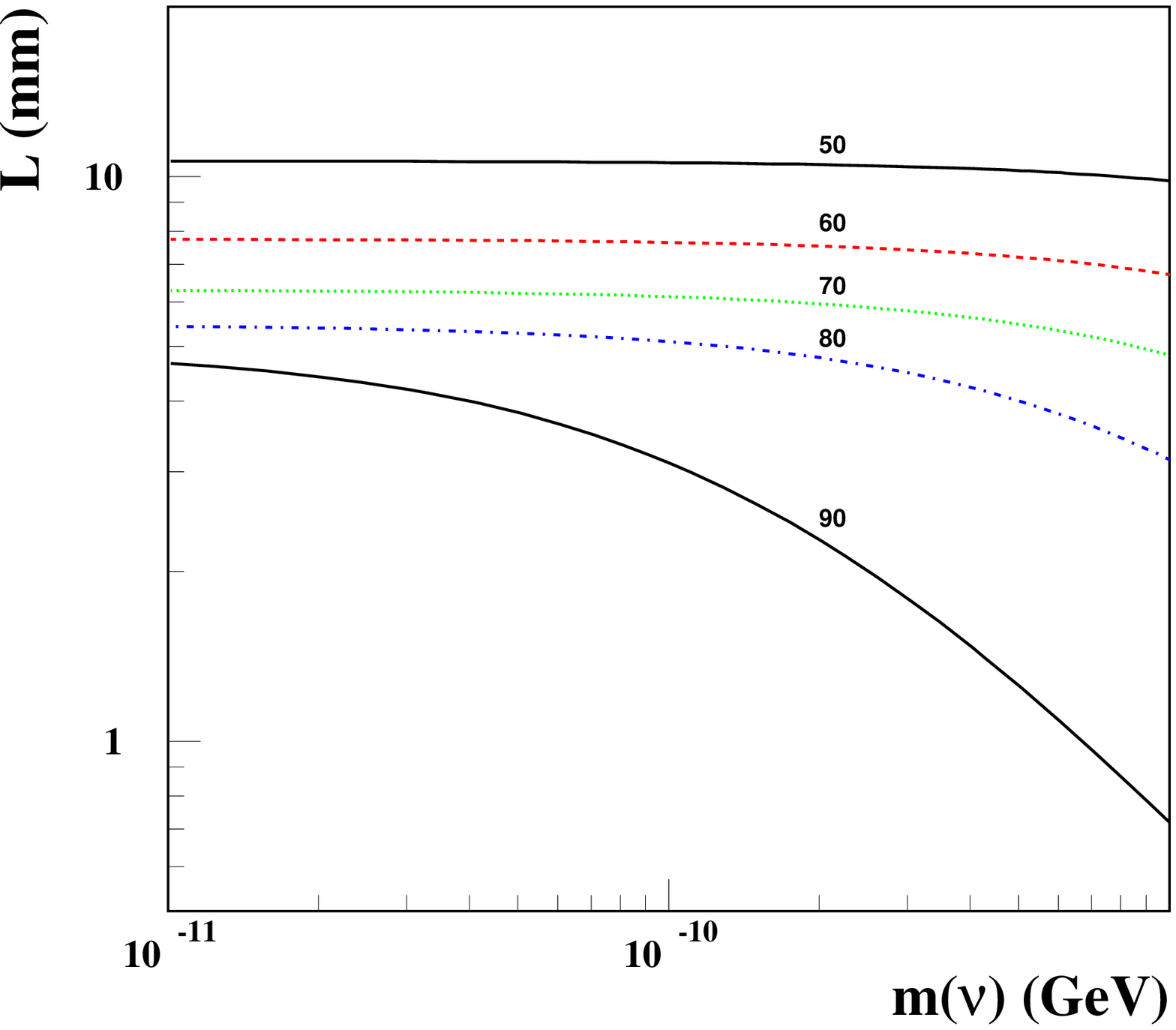}
 \caption{  $\tilde{\chi}_1^0$  decay length versus LSP mass for 
   $A_0=0$, $\mu > 0$ and $\tan\beta=3$ for (a) fixed BRpV parameters:
   $\epsilon_\tau = 0.22$ GeV, $m_{\nu_3} = 0.05$ eV (solid lines) and current
   atmospheric 3$\sigma$ band (shaded bands); (b) as a function of the
   neutrino mass $m_{\nu_3}$ for $m_0=100$ GeV and several values of
   $\tilde{\chi}_1^0$ masses .}
\label{fig:dpath}
\end{figure*}

Fig.\ \ref{fig:dpath}(a) shows the lightest neutralino decay length as a
function of its mass for $A_0 = 0$, $\mu > 0$, $\tan\beta = 3$, and
$\epsilon_\tau = 0.22$ GeV.  In this figure, the solid lines stand for the
tree-level neutrino mass $m_{\nu_3}=0.05$ eV, corresponding to the best fit
atmospheric scale as given in Ref.~\cite{Maltoni:2002ni}, and we chose $m_0 =
200$ and 700 GeV. The bands in these figures were obtained by taking
$m_{\nu_3}$ within the 3$\sigma$ allowed atmospheric mass splitting of
Ref.~\cite{Maltoni:2002ni}.  Fig.\ \ref{fig:dpath}(b) presents the LSP decay
length as a function of the neutrino mass, for different lightest neutralino
masses and the parameters used in Fig.\ \ref{fig:dpath}(a).

From Figs.\ \ref{fig:dpath} we can see clearly that the LSP decay length is
shorter for larger neutrino and LSP masses, as expected.  Furthermore,
irrespective of the smallness of the neutrino mass indicated by the
atmospheric oscillation data, the LSP ($\tilde{\chi}^0_1$) decays inside the
detector in a large portion of the parameter space, specially for small $m_0$.
It is interesting to notice that the kinks in Fig.\ \ref{fig:dpath} are
associated to the opening of the LSP decays into on-shell gauge bosons.

\begin{figure*}[thbp] 
\includegraphics[height=9cm,width=0.46\linewidth]{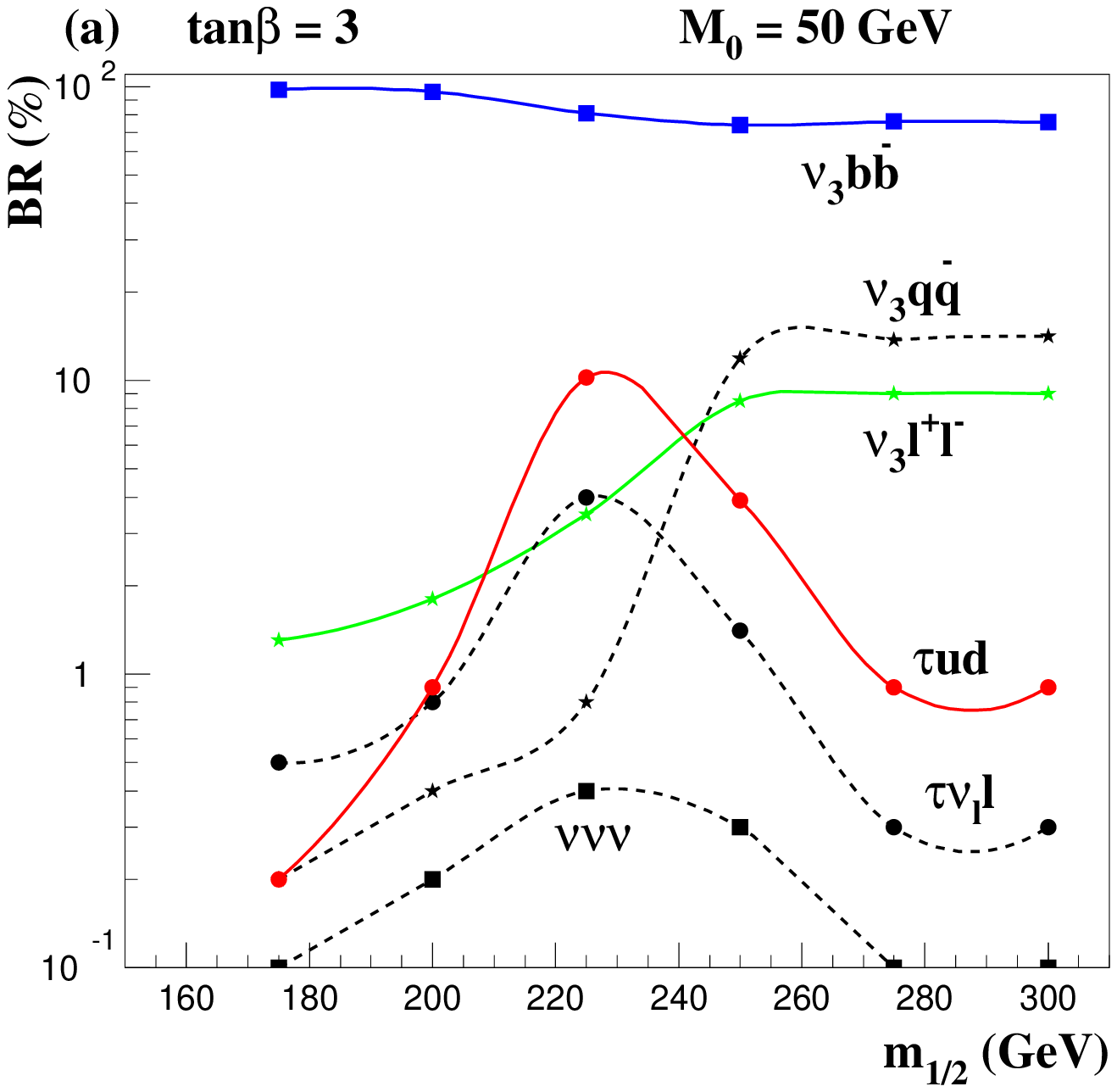}
\includegraphics[height=9cm,width=0.46\linewidth]{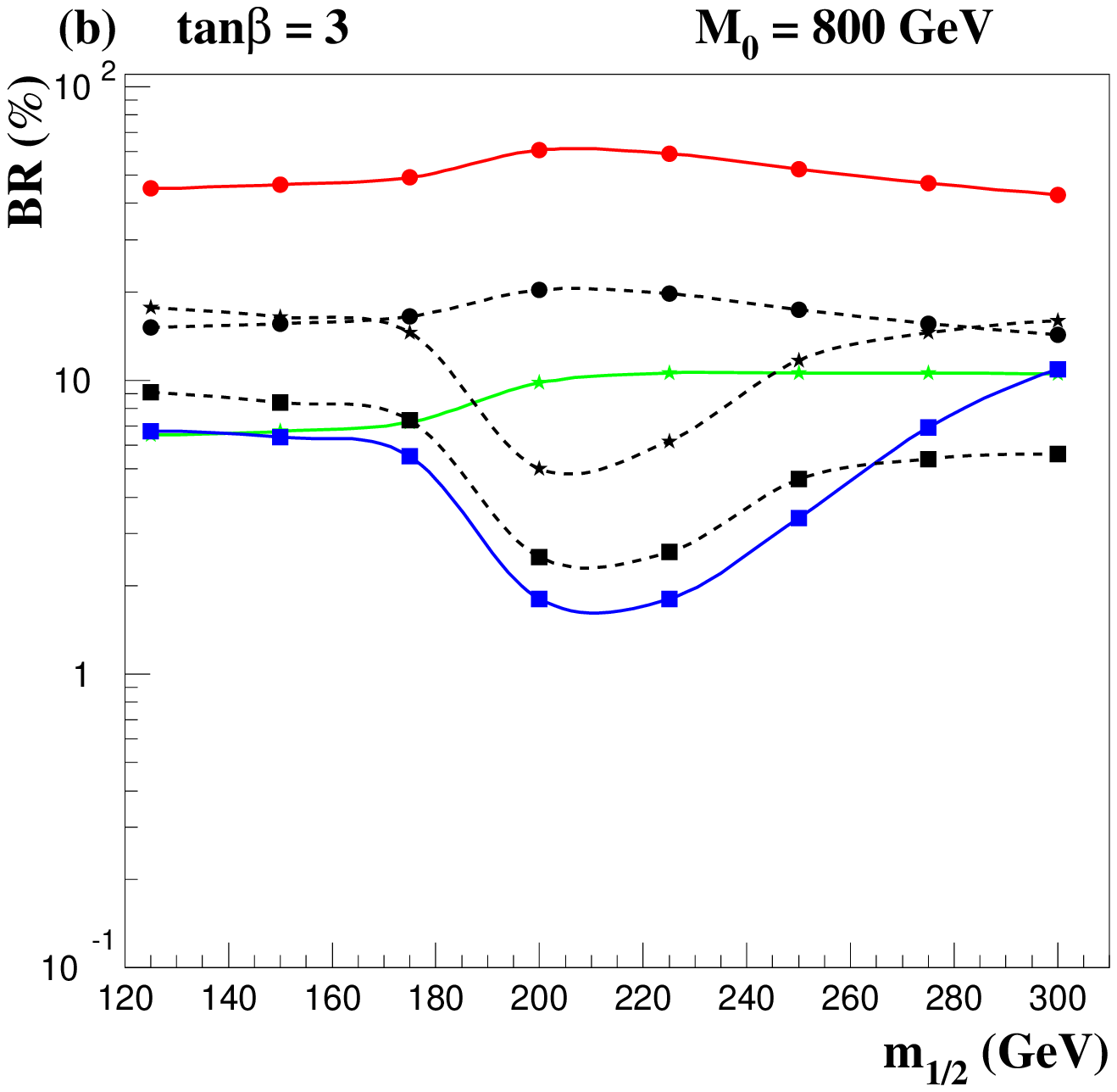}
\includegraphics[height=9cm,width=0.46\linewidth]{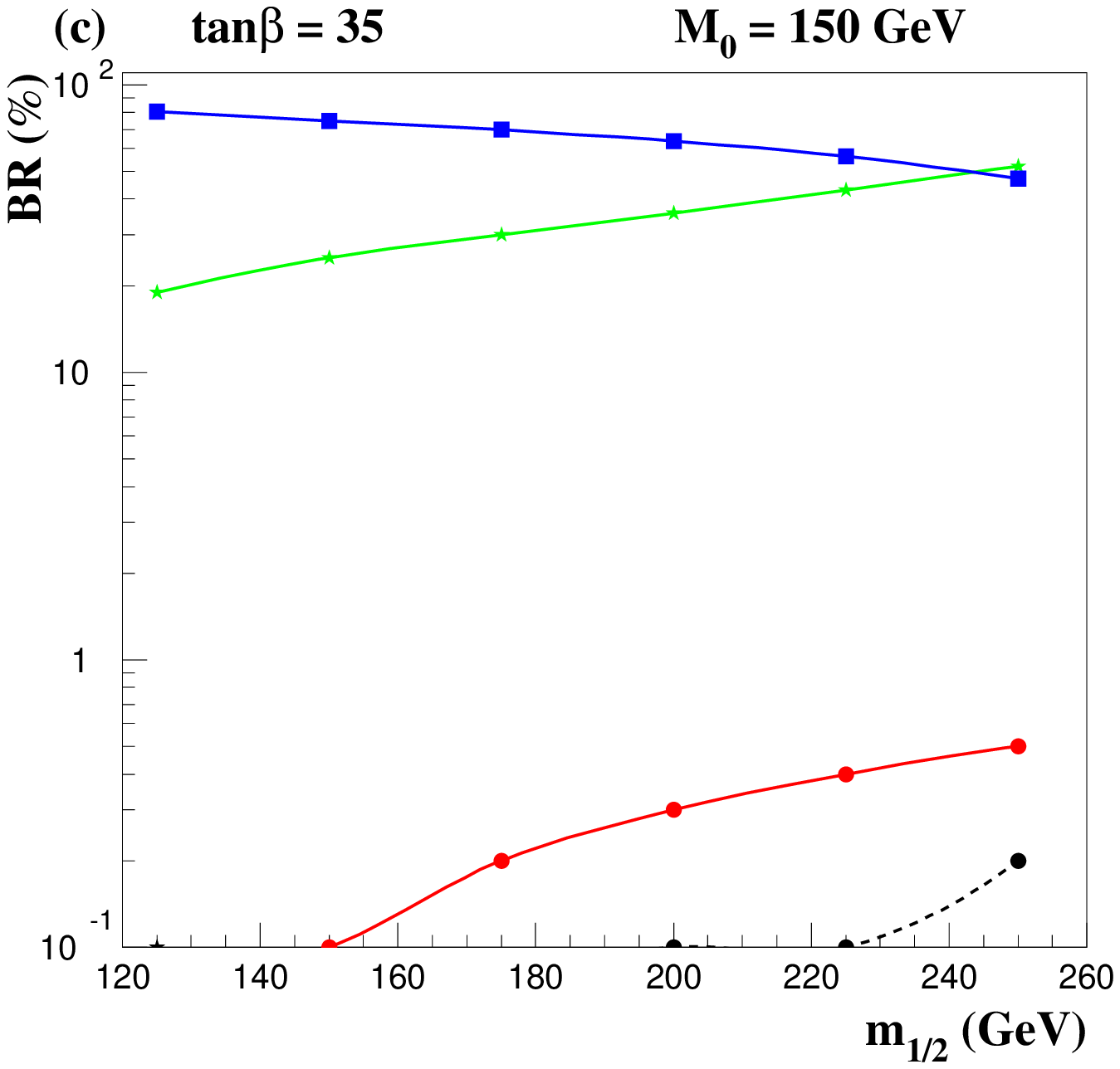}
\includegraphics[height=9cm,width=0.46\linewidth]{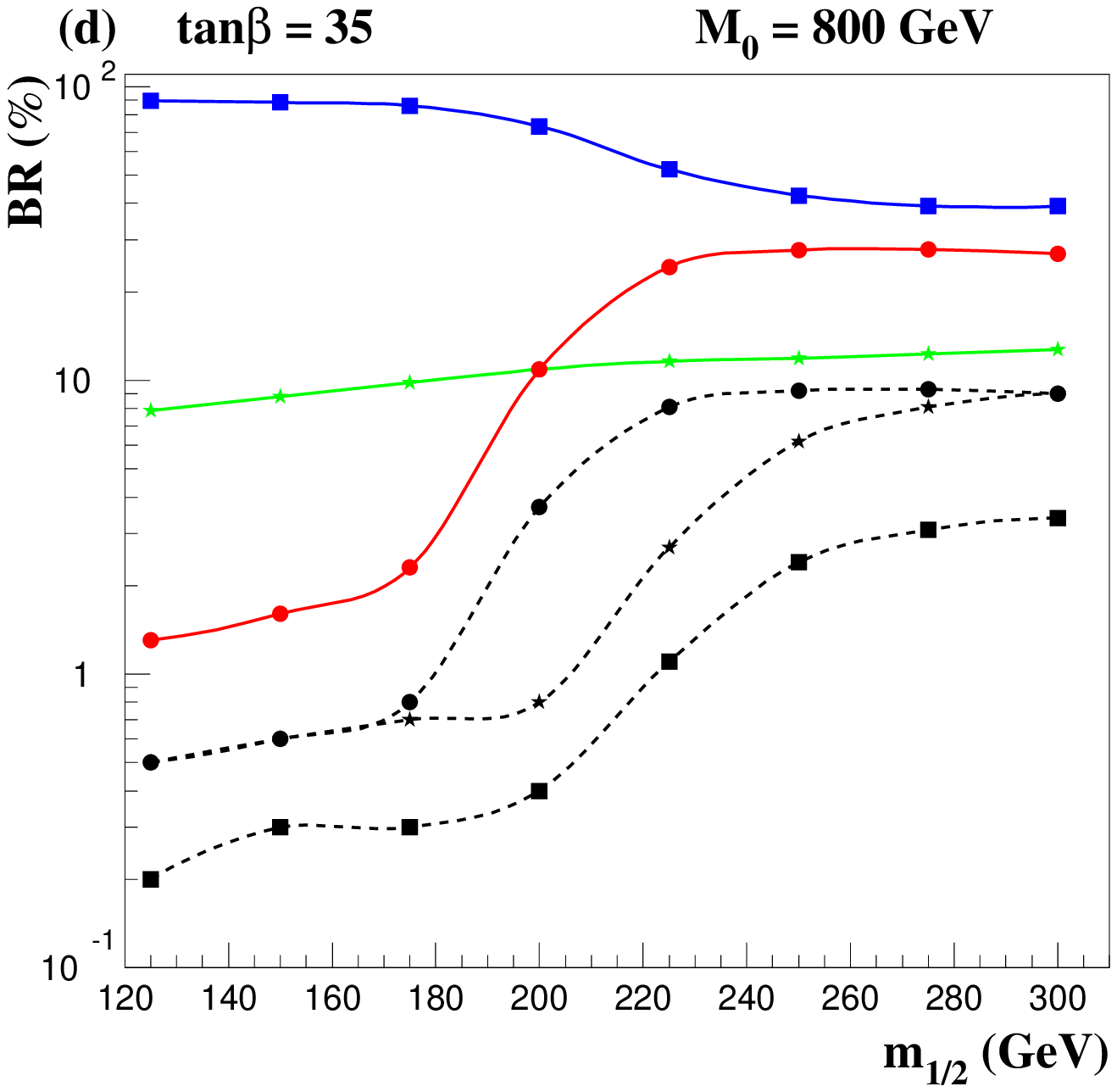}
\caption{ $\tilde{\chi}_1^0$ branching ratios as a function of
  $m_{1/2}$ for $A_0=0$, $\mu > 0$, $\epsilon_\tau = 0.22$ GeV, and
  $m_{\nu_3} = 0.05$ eV. The solid lines denote $\tilde{\chi}^0_1 \to
  \nu_3 b \bar{b}$ (squares); $\tilde{\chi}^0_1 \to \tau u \bar d$ 
  (circles); and $\tilde{\chi}^0_1 \to \nu_3 \ell^+ \ell^-$
  (stars). The dashed lines denote $\tilde{\chi}^0_1 \to $ invisible
  (squares); $\tilde{\chi}^0_1 \to \tau \nu_{\ell}
  \ell$ (circles); and
  $\tilde{\chi}^0_1 \to \nu_3 q \bar{q}$ (stars) .}
\label{fig:1}
\end{figure*}

Figure \ref{fig:1} contains the $\tilde{\chi}^0_1$ branching ratios as a
function of $m_{1/2}$ for $\epsilon_\tau = 0.22$ GeV, $m_{\nu_3} = 0.05$ eV,
$A_0=0$, and $\mu > 0$.  As we can see from this figure, the decay
$\tilde{\chi}^0_1 \to \nu b \bar{b}$ dominates at small $m_0$ and large $\tan
\beta$. This decay channel arises from an effective coupling
$\lambda^\prime_{333}$, induced by the neutrino--neutralino mixing, which is
proportional both to the bottom Yukawa coupling and to the bilinear parameter
$\epsilon_\tau$ in our model.  Therefore, this decay is enhanced at small
$m_0$ due to the lightness of the scalars that mediate it and at large
$\tan\beta$ due to enhanced Yukawa couplings.  Both features are apparent in
the panels of this figure.  Moreover, whenever $\tilde{\chi}^0_1 \to \nu b
\bar{b}$ is not the leading channel, the LSP decays mostly to $ \tau u \bar d
$, with a sizable branching fraction $\tilde{\chi}^0_1 \to \tau \nu_\ell \bar
\ell$.

\begin{figure*}[thbp] 
\includegraphics[height=9cm,width=0.46\linewidth]{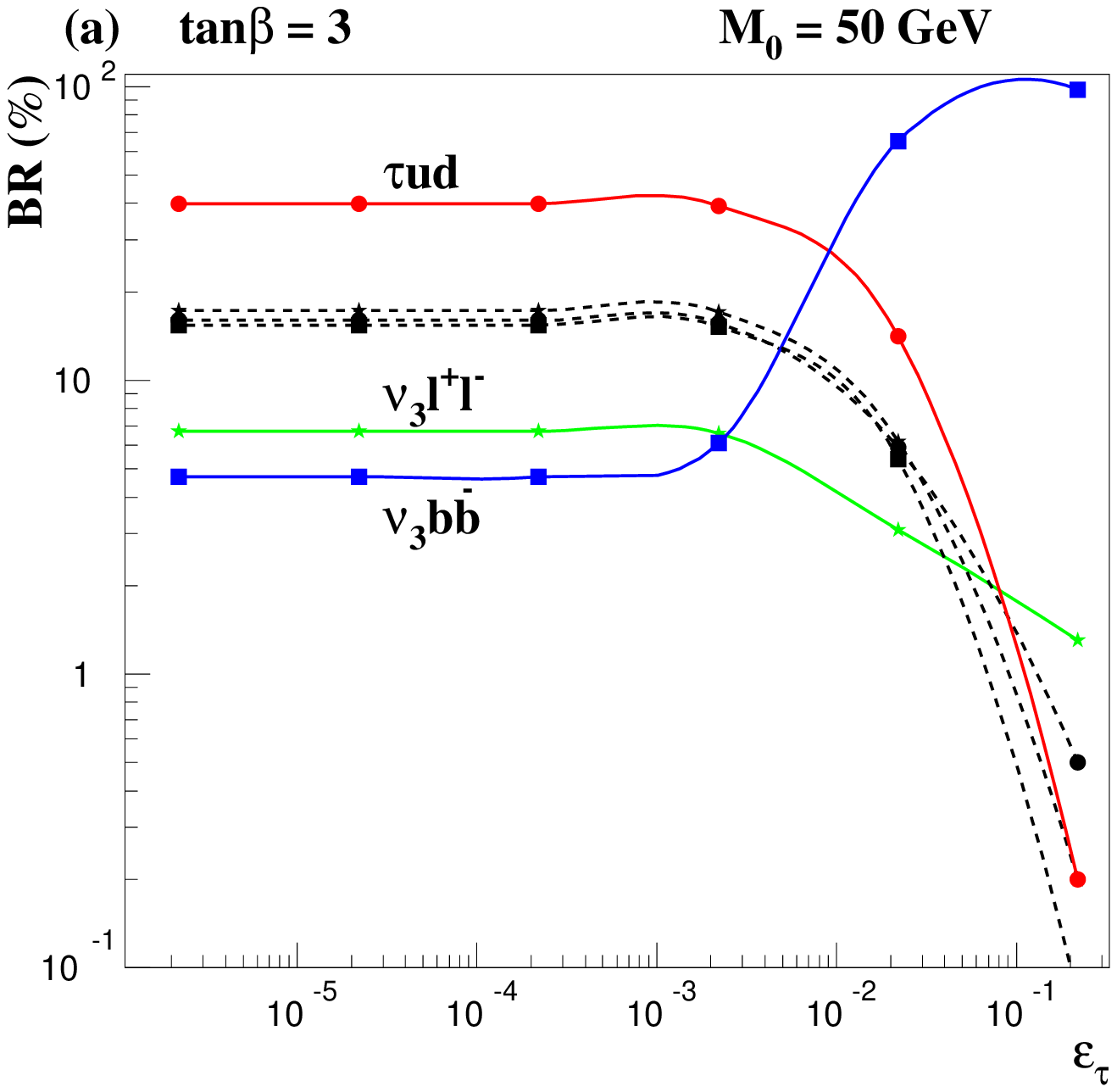}
\includegraphics[height=9cm,width=0.46\linewidth]{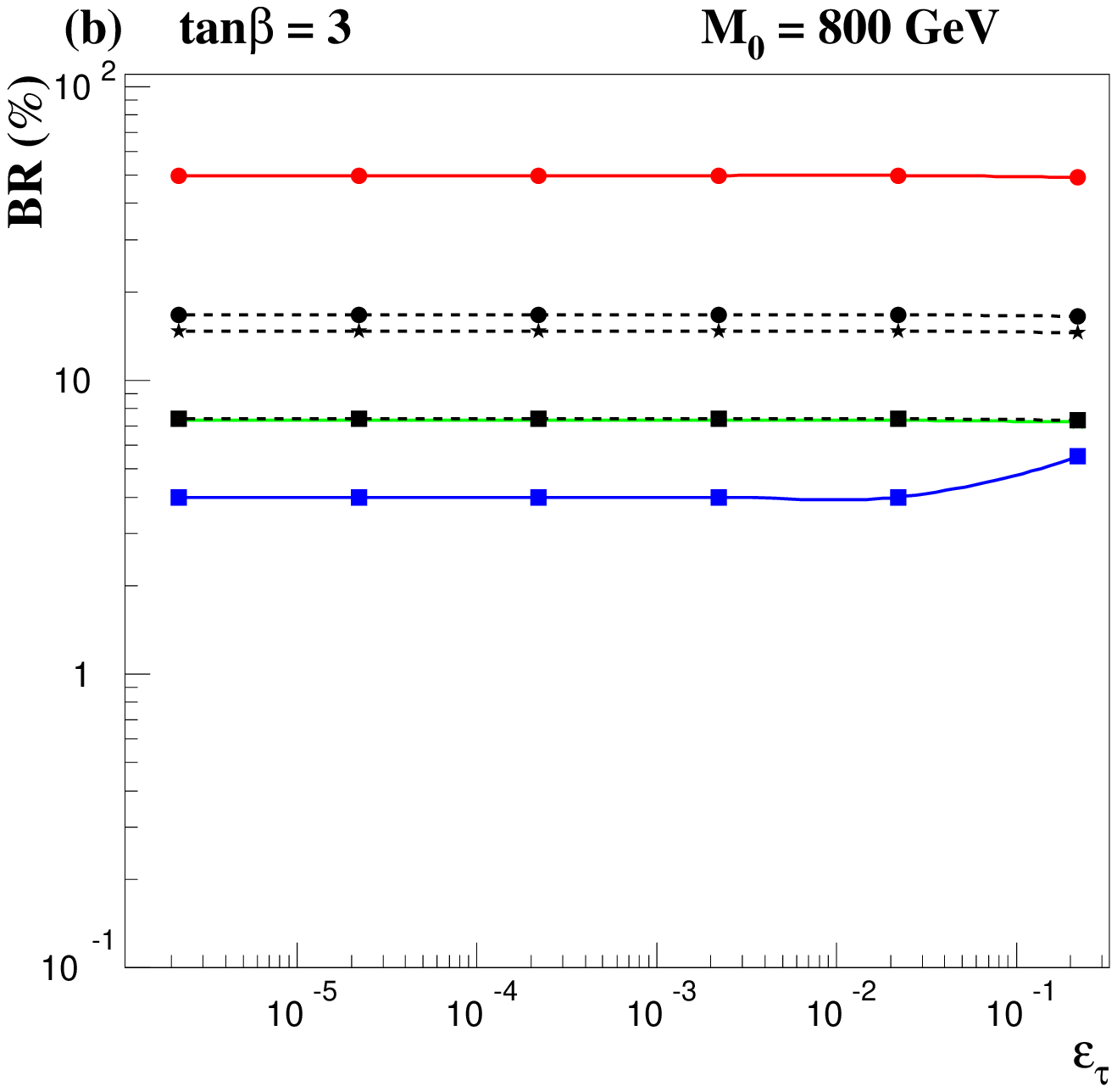}
\includegraphics[height=9cm,width=0.46\linewidth]{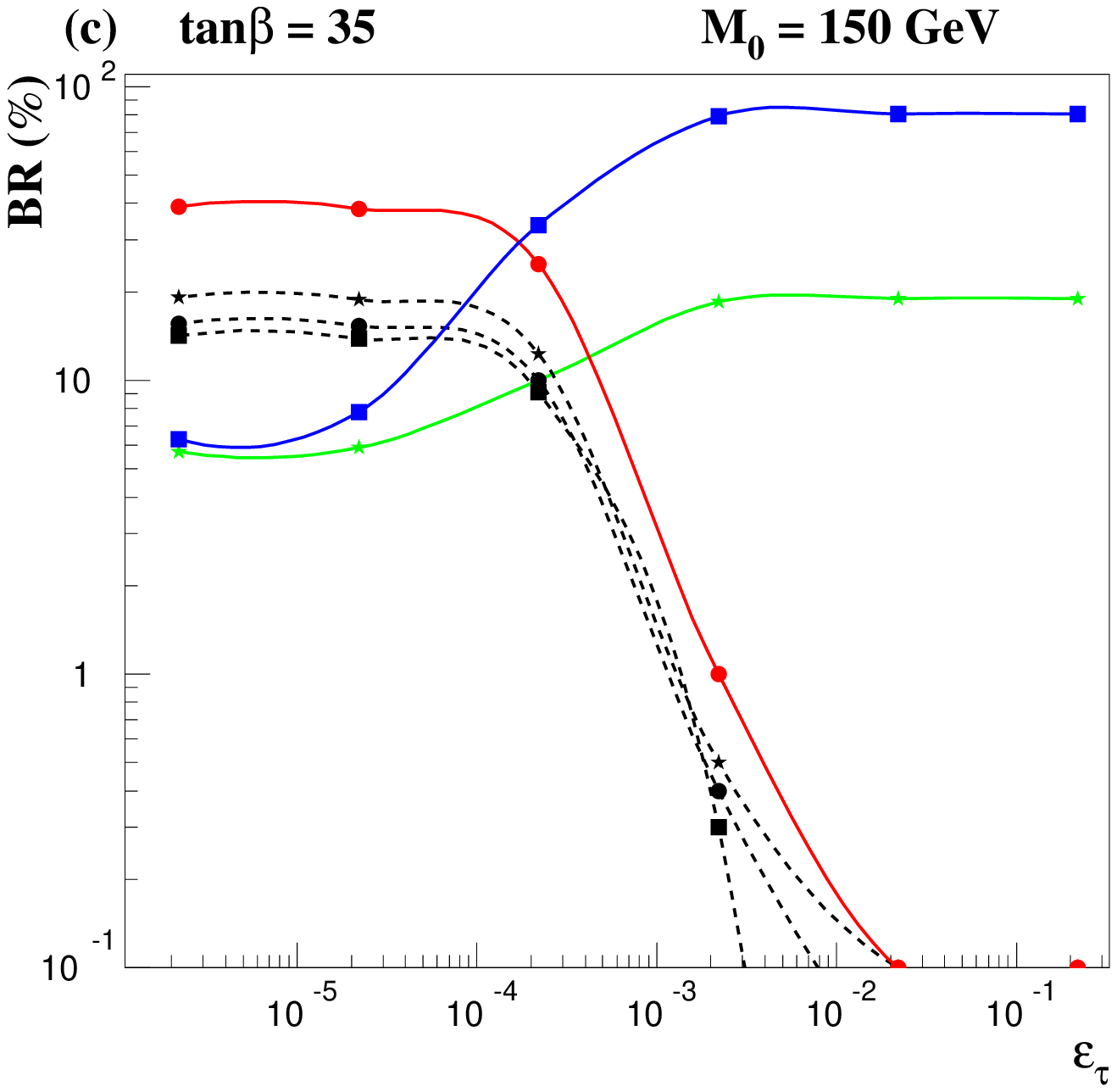}
\includegraphics[height=9cm,width=0.46\linewidth]{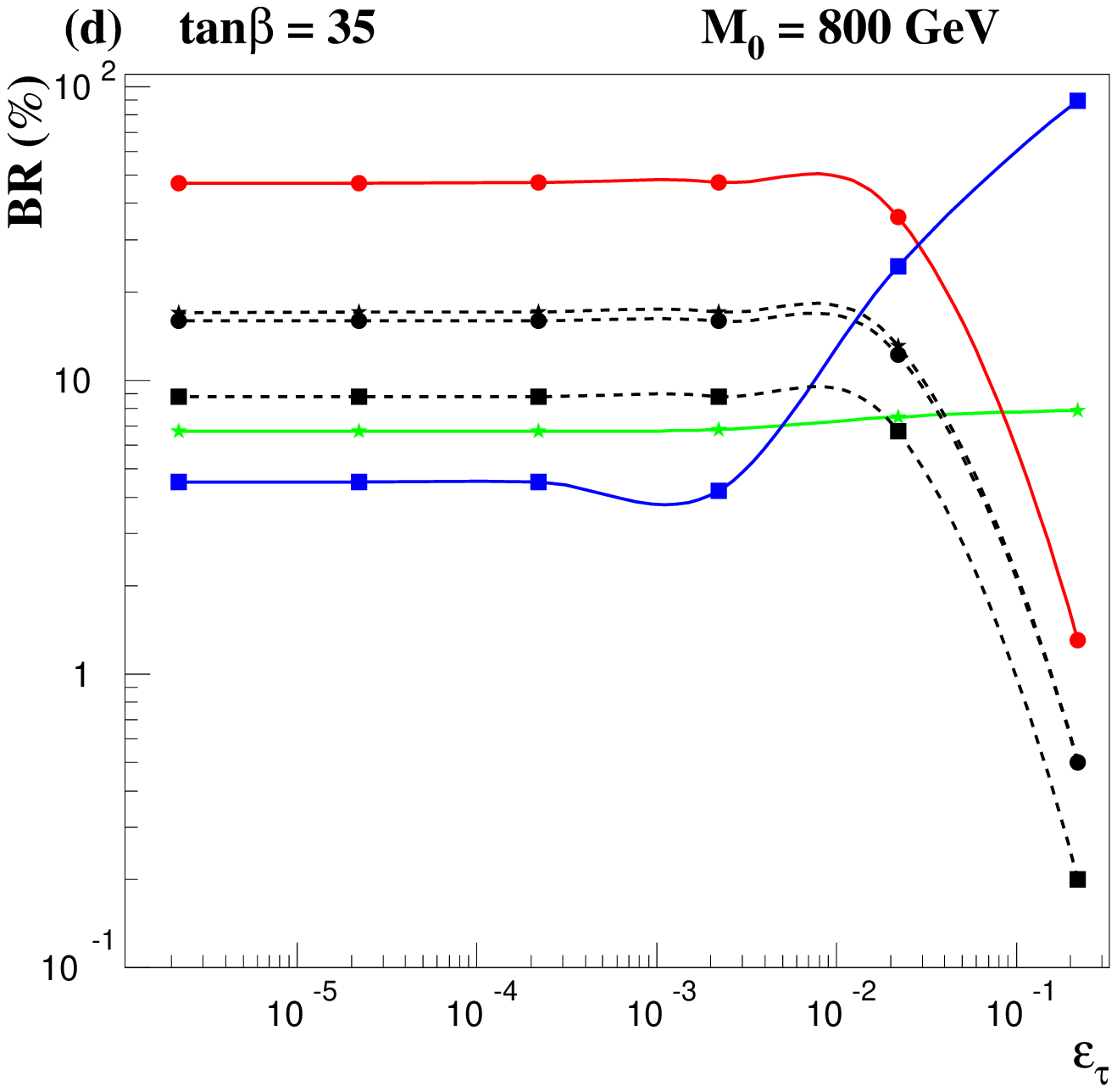}
\caption{ $\tilde{\chi}_1^0$ branching ratios as a function of
  $\epsilon_\tau$ for $A_0 =0$, $\mu > 0$, and $m_{\nu_3} = 0.05$ eV.
  We fixed $m_{1/2} = 175$ GeV in the case of $\tan\beta = 3$ and
  $m_{1/2} = 125$ GeV for $\tan\beta = 35$. The lines are as in Fig.\
  \ref{fig:1}.  }
\label{fig:1.1}
\end{figure*}

In Fig.\ \ref{fig:1.1}, we present the dependence of the $\tilde{\chi}^0_1$
branching ratios with respect to the $R$-parity violating parameter
$\epsilon_\tau$. The importance of the $\nu b \bar{b}$ decay mode increases
for large $\epsilon_\tau$, since the effective coupling $\lambda^\prime_{333}$
is proportional to $\epsilon_\tau$.  Moreover, Fig.\ \ref{fig:1.2} shows that
the LSP branching ratio into $\nu b \bar{b}$ decreases with increasing
$m_{\nu_3}$ in the parameter regions where this mode is sizeable.

One comment is in order here, concerning the naturalness of the above
branching ratio predictions. This issue depends somewhat on the
assumptions about the supersymmetry soft breaking terms. We are
tacitly assuming them to be universal at some unification scale. This
is the usual practice and is adopted here to ensure simple comparison
with the SUGRA $R$-parity conserving results for the trilepton signal
obtained in Ref.~\cite{tata1}.
In the BRpV model universality plays another important role, namely,
it ensures ``calculability'' of the neutrino mass $m_{\nu_3}$, by the
renormalization group evolution. One finds in this case that the
smallness of $m_{\nu_3}$ required by experiment follows naturally from
the approximate ``alignment'' of BRpV parameters discussed, for
example, in Ref.~\cite{BRpB_tau,Mira:2000gg}. In such scenario one has
that the short LSP decay path is indeed technically natural, despite
the small neutrino mass.  However one should stress that strictly
universal boundary conditions are not at all essential for the
consistency of the results presented here.

Summing up, we can say that in the BRpV model the $\tilde{\chi}_1^0$ decays
mainly into $\tau u d$ for large $m_0$ or small $\epsilon_\tau$, while its
decays are dominated by $\nu b \bar{b}$ at small $m_0$ and large
$\epsilon_\tau$ ($\tan\beta$).  As we will see in what follows this has
important implications for the trilepton signal.

%%%%%%%%%%%%%%%%%%%%%%%%%%%%%%%%%%%%%%%%%%%%%%%%%%%%%%%%%%%%%%%%%%%%%%
\section{Signal, Backgrounds, and Selection Cuts}

In $R$-parity conserving scenarios, trilepton production at the Tevatron
proceeds via $ p \bar{p} \to \tilde{\chi}^0_2 \tilde{\chi}^\pm_1$ with
$\tilde{\chi}^\pm_1 \to \ell \nu \tilde{\chi}^0_1$, $\tilde{\chi}^0_2 \to
\ell^+ \ell^- \tilde{\chi}^0_1$, and the LSP ($\tilde{\chi}^0_1$) leaving the
detector undetected. The main SM backgrounds for trilepton production are
displayed in Table \ref{table1}.  In order to suppress these backgrounds, we
have imposed the soft cuts SC2 defined in Ref.\ \cite{tata1}, which were
tailored for scenarios containing soft signal leptons coming from $\tau$
decays \cite{chung}:

\begin{itemize}
 
\item[C1:] We required the presence of three isolated leptons ($e$ or
    $\mu$) with a hadronic $E_T$ smaller that 2 GeV in a cone of size
    $\Delta R = 0.4$ around the lepton \cite{3l-4l};
  
\item[C2:] We required the most energetic lepton to satisfy $| \eta_{\ell_1} |
  < 1.0$ and $p_T(\ell_1) > 11$ GeV;
    
\item[C3:] The second (third) most energetic lepton must satisfy $|
  \eta_{\ell_{2/3}} | < 2.0$ and $p_T(\ell_{2(3)}) > 7$ (5) GeV;
    
\item[C4:] We required the missing transverse energy to be larger than 25 GeV;
  
\item[C5:] We vetoed events exhibiting a $\ell^+ \ell^-$ pair with an
  invariant mass smaller than 20 GeV and larger than 81 GeV (this
  avoids both Z boson and QED contributions);
  
\item[C6:] We vetoed events with a transverse mass of a charged lepton
    and missing transverse energy between 60 GeV and 85 GeV in order to
    suppress $W$ decays.

\end{itemize}

In our analysis, the signal and backgrounds were generated using PYTHIA
\cite{pythia}, except for the $W Z^\star (\gamma^\star )$ which was computed
using the complete matrix elements \cite{madgraph}.  Furthermore, we also
simulate experimental resolutions by smearing the energies, but not
directions, of all final state particles with a Gaussian error given by
$\Delta E/E = 0.7/\sqrt{E}$ ($E$ in GeV) for hadrons and $\Delta E /E =
0.15/\sqrt{E}$ for charged leptons.

The cross section for the SM backgrounds after cuts are shown in Table
\ref{table1}. Notice that our results differ slightly from the ones in Ref.\ 
\cite{tata1} because the hadronization procedures used in PYTHIA and ISAJET
are different. Moreover, we can see that the most important background is the
$t \bar{t}$ production, accounting for 70\% of the total background. We
further tested our code by verifying that our results for the $R$-parity
conserving signal agree with the ones in Ref.\ \cite{tata1}.

%%%aqui

In our BRpV model there are more SUSY reactions that can contribute to the
trilepton signal than in the $R$-parity conserving case, since the
$\tilde{\chi}_1^0$ decays can give rise to charged leptons.  In the parameter
range interesting for neutrino physics $R$-parity violating decay modes are
only important for the LSP decays and can safely be neglected for the other
SUSY particles.  We have incorporated all possible decay modes of the lightest
neutralino and the second lightest neutralino taking fully into account large
$\tan(\beta)$ effects in PYTHIA, leaving the other decay modes unchanged.
Assuming that gluinos and squarks are too heavy to be produced at the
Tevatron, we considered the following processes:
\[
p \bar{p} \to \tilde{\ell} \tilde{\ell}^\star \;\;\;,\;\;\; 
\tilde{\nu} \tilde{\ell} \;\;\; , \;\;\;
\tilde{\chi}^0_i \tilde{\chi}^0_j~ (i(j)=1,2) \;\;\;,\;\;\; 
\tilde{\chi}^+_1 \tilde{\chi}^-_1 \;\;\;\hbox{,  and} \;\;\;
\tilde{\chi}^0_i \tilde{\chi}^\pm_1 (i=1,2) \;.
\]

The $\tilde{\chi}^0_1$ decays can contain charged leptons, and therefore, we
should also analyse multilepton ($\ge 4 \ell$) production. In order to
extract this signal, we applied the cuts C1, C3, C5, and C2 but accepting
leptons with $|\eta(\ell)| < 3$. We also required the presence of an
additional isolated lepton with $p_T > 5$ GeV and demanded the missing
transverse energy to be larger than 20 GeV. The main SM backgrounds for this
process are $t \bar{t}$, $WZ$, and $ZZ$ productions whose cross sections after
cuts are shown in Table \ref{table2}.

We can see from Fig.\ \ref{fig:dpath} that the lightest neutralino might not
decay inside the detector depending on the point of the parameter space. If it
decays inside the tracking system, it can give rise to spectacular events
exhibiting displaced vertices without an incoming track associated to it.  In
our analyses, we did not look for displaced vertices since the corresponding
backgrounds depend upon details of the detector. This possibly is a
conservative hypothesis since this kind of events should present a small
background, leading to a larger reach of the Tevatron.  Nevertheless, we kept
track of the position of the neutralino decay and accepted events where the
neutralino decays inside the tracking system, rejecting all events where one
of the neutralinos decays outside a cylinder around the beam pipe of radius
0.5 m and length 2.0 m; we named this requirement C7. Again, this is another
conservative estimate of the expected BRpV signal.

We investigated the regions of the $m_0 \otimes m_{1/2}$ plane where the
trilepton and multilepton signals can be established at the Tevatron for
integrated luminosities of 2 fb$^{-1}$ and 25 fb$^{-1}$ and fixed values of
$A_0$ (=0), $\tan\beta$, ${\mathrm{sign}}(\mu)~(>0)$, $\epsilon_\tau$, and
$m_{\nu_3}$ (= 0.05 eV).  We evolved the renormalization group equations
starting at the unification scale (few $\times 10^{16}$ GeV) for the soft
parameters, rejecting the points where either the electroweak symmetry is not
correctly broken, or which exhibit particles with mass excluded by present
experimental data~\cite{pdg}.  In our analysis, we have employed the Poisson
statistics, except when the expected number of signal events is large enough
to justify the use of the Gaussian distribution \cite{tatadif}.  We exhibit
our results in the $m_0 \otimes m_{1/2}$ plane, denoting by black circles the
theoretically excluded points, and by white circles the experimentally
excluded by sparticle and Higgs boson searches at LEP2~\cite{pdg}. The black
squares represent points accessible to Tevatron experiments at $5\sigma$ level
with 2 fb$^{-1}$ of integrated luminosity, while the white squares are
accessible with 25 fb$^{-1}$.  Points denoted by diamonds are accessible only
at the $3\sigma$ level with 25 fb$^{-1}$, while the stars correspond to the
region not accessible to the Tevatron.

The trilepton cross section is always dominated by the $\tilde{\chi}^+_1
\tilde{\chi}^-_1$ and $\tilde{\chi}^0_2 \tilde{\chi}^\pm_1$ productions, with
the second process being about 20\% to 50\% larger than the first.  For
example, these reactions are responsible for approximately 99\% of the cross
section at large $m_0$. Note that the first process can contribute to the
trilepton signal only in $R$-parity violating scenarios.  Moreover, at small
and moderate $m_0$ ($\lesssim 400$ GeV), the $\tilde{\chi}^0_1
\tilde{\chi}^\pm_1$ production is responsible for approximately 5--10\% of the
signal cross section and the production of sleptons also gives a sizable
contribution.

To study the multilepton signal we choose three representative
parameter regions.  The best scenario corresponds to the case where
$\tan\beta$ and $\epsilon_\tau$ are small. We subsequently relax this
optimistic assumption by considering separately the cases where either
(but not both, since in this case the signal is too small)
$\epsilon_\tau$ or $\tan\beta$ are large.

\begin{figure}
\begin{center}
\parbox[l]{3.5in}{
\mbox{\includegraphics[height=9cm,width=0.96\linewidth]{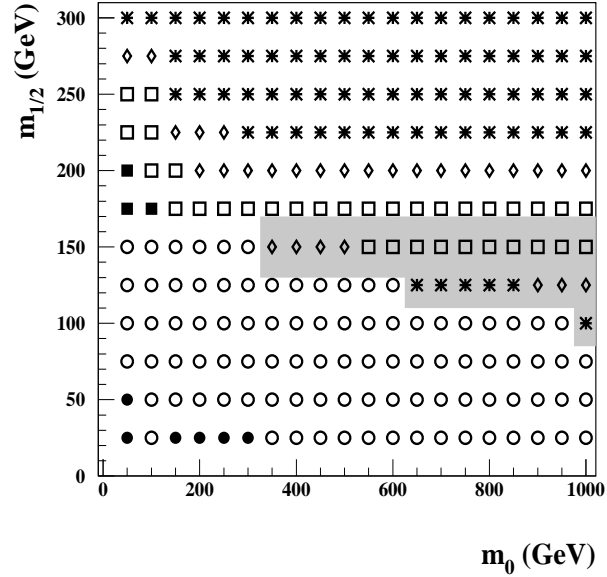}}
}
\hfill
\end{center}
\caption{
  Reach of Fermilab Tevatron Run II using the trilepton signal in the $m_0
  \otimes m_{1/2}$ plane for $A_0=0$, $\tan\beta = 3$, $\mu > 0$,
  $\epsilon_\tau = 7\times 10^{-4}$ GeV, and $m_{\nu_3} = 0.05$ eV.  The black
  circles are theoretically excluded, while the white circles are
  experimentally excluded by sparticle and Higgs boson searches at LEP2. The
  black squares denote points accessible to Tevatron experiments at $5\sigma$
  level with 2 fb$^{-1}$ of integrated luminosity, while the white squares are
  accessible with 25 fb$^{-1}$. Points denoted by diamonds are accessible at
  the $3\sigma$ level with 25 fb$^{-1}$, while the stars correspond to the
  region not accessible to Tevatron.  The long lifetime of the neutralino
  reduces considerably the signal in the shaded area, however, it suggests
  that the sensitivity can be improved by looking for displaced vertices.
}
\label{fig:2}
\end{figure}

In Fig.\ \ref{fig:2}, we present the region of the $m_0 \otimes m_{1/2}$ plane
that can be probed at the Tevatron for $\tan\beta = 3$ and BRpV parameters
$\epsilon_\tau = 7 \times 10^{-4}$ GeV and $m_{\nu_3} = 0.05$ eV.  
For these values of the parameters,
the signal cross section is dominated by $\tilde\chi_1^+\tilde\chi_2^0 $
production followed by $\tilde\chi_1^+\to\tilde\chi_1^0l^+\nu$,
$\tilde\chi_2^0\to\tilde\chi_1^0qq$ and the LSP $\tilde\chi_1^0$ decays mainly
into $\tau u \bar d$. As expected, the long lifetime of the neutralino reduces
considerably the signal in the shaded area of Fig.\ \ref{fig:2} after we apply
the cut C7. Therefore, we might be able to further probe this region by
looking for displaced vertices, and consequently enhance the Tevatron reach.

%%%% eps small tb small %%%%%%%%%%%%%%%

It is interesting to compare our results presented in Fig.\ \ref{fig:2} with
the ones in Ref.\ \cite{tata1}.  First of all, the presence of BRpV
interactions reduces the Tevatron reach in the trilepton channel for small
values of $m_0$.  This happens because there are some competing effects in
this region of parameters: on the one hand there are new contributions to the
trilepton process due to LSP decay and on the other hand, the decay of the
neutralinos produce a larger hadronic activity, worsening the lepton
isolation, and reducing the missing $E_T$ compared with the MSSM case.
Besides that, the leptons from the $\tilde{\chi}^0_1$ decay can give rise to
additional isolated leptons which can contribute to the trilepton signal or,
alternatively, can suppress it due to the presence of more than three isolated
leptons.  The last effect and the larger hadronic activity reduce the
trilepton signal at small $m_0$ in the BRpV model.  In contrast, as can be
seen from Fig.\ \ref{fig:2}, the trilepton reach at large $m_0$ always tends
to increase with respect to the MSSM expectation.  This follows from the fact
that the drastic reduction of the $\tilde{\chi}^0_2$ branching ratio into
leptons at large $m_0$ is compensated by the additional production of charged
leptons in $\tilde{\chi}_1^0$ decays.  Since these extra leptons come from tau
decays, it is important to adopt cuts which increase the acceptance of soft
leptons.  The cuts we used satisfy this requirement.

\begin{figure}
\begin{center}
\parbox[l]{3.5in}{
\mbox{\includegraphics[height=9cm,width=0.96\linewidth]{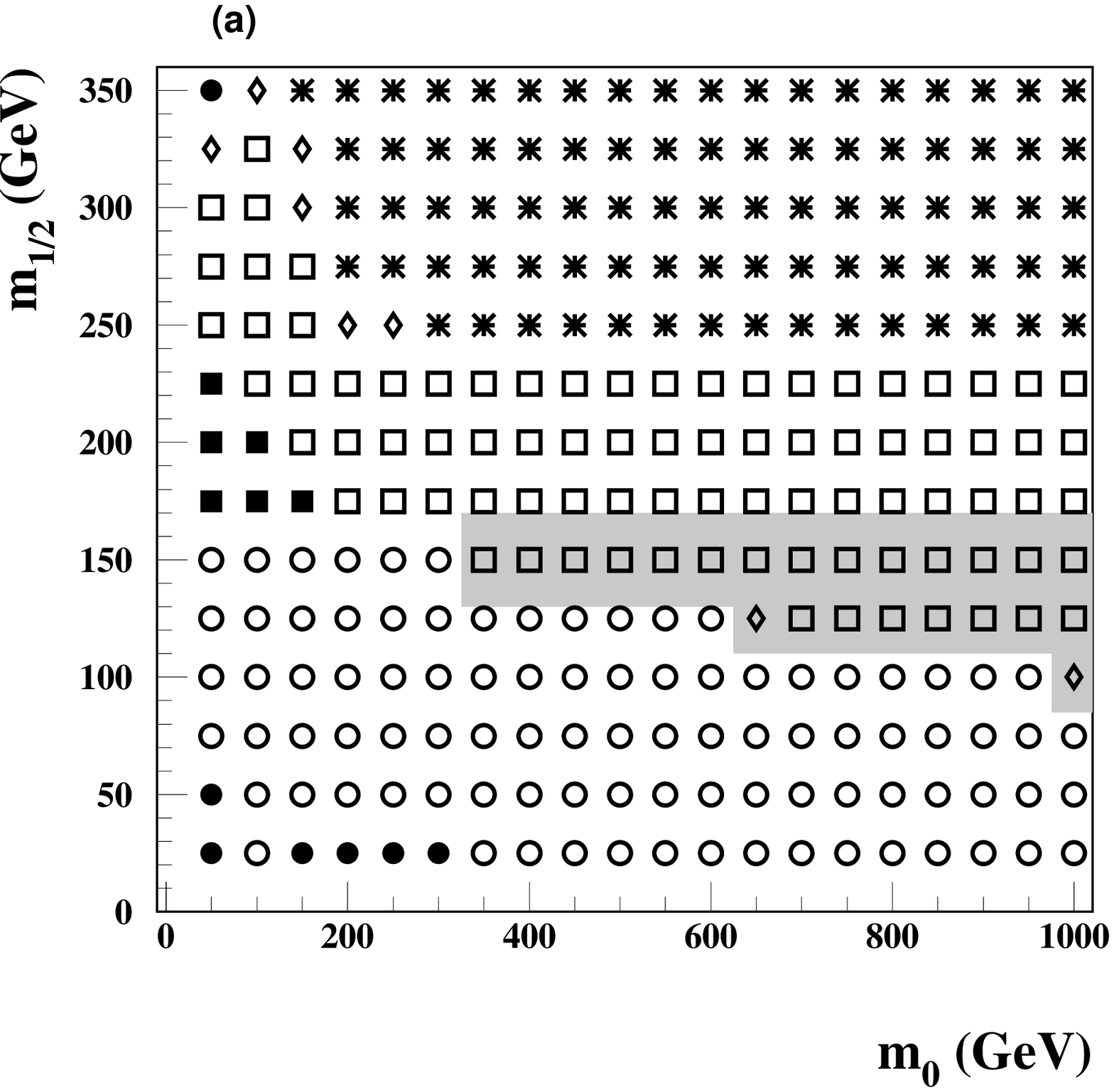}}
}
\hfill
\parbox[l]{3.5in}{
\mbox{\includegraphics[height=9cm,width=0.96\linewidth]{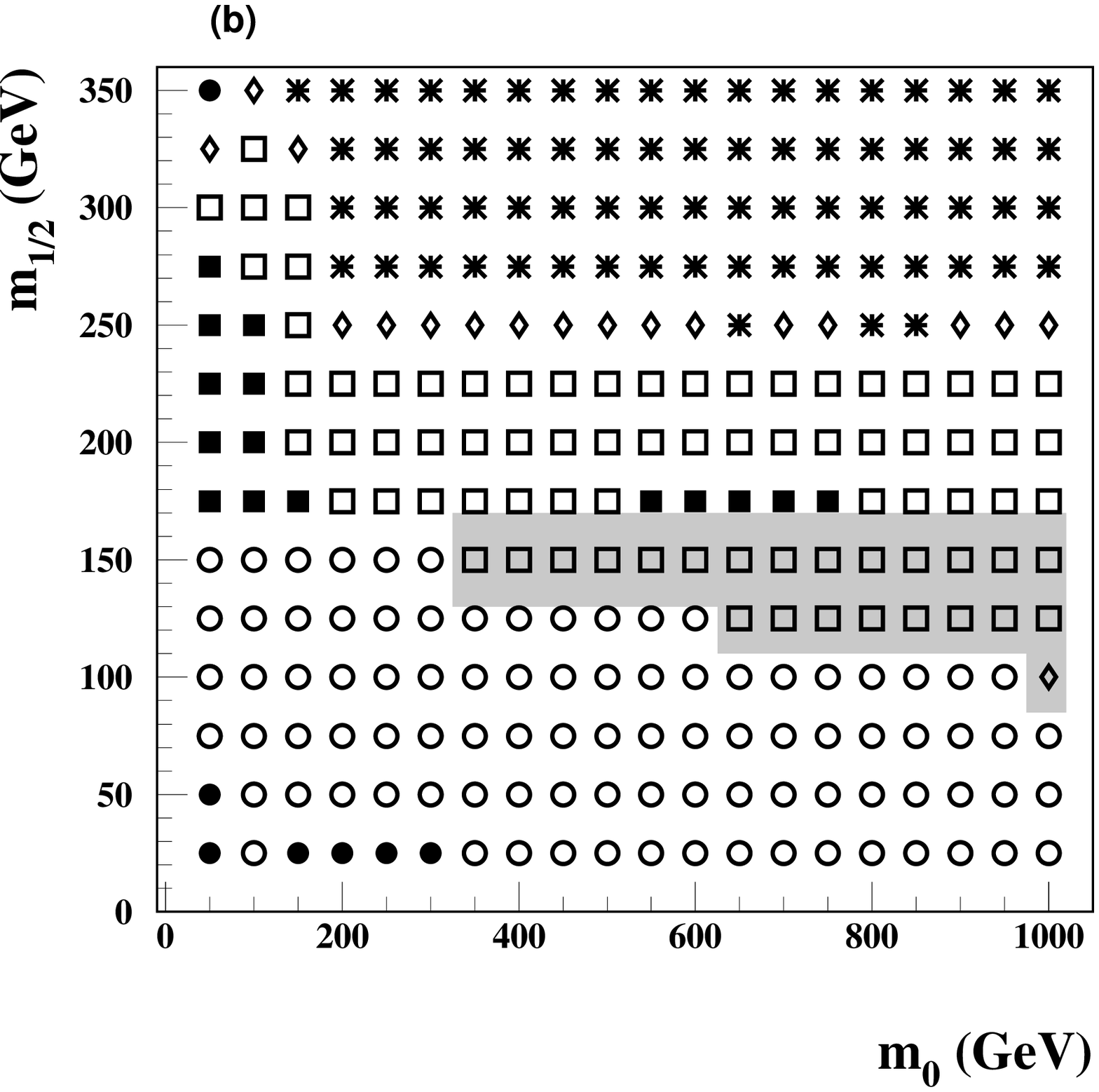}}
}  
\end{center}
\caption{
        (a) Reach of Fermilab Tevatron Run II in the 4 or more lepton
        channel. (b) Combined trilepton and multilepton results. All
        parameters and conventions were chosen as in Fig.\ \ref{fig:2}.
}
\label{fig:3}
\end{figure}

In Fig.\ \ref{fig:3}(a) we present the Tevatron reach in the multilepton
(four leptons or more) channel for the same parameters adopted in Fig.\ 
\ref{fig:2}. As we can see, the multilepton reach is larger than the
trilepton one, increasing the discovery potential at large values of
$m_{1/2}$. For instance, the Tevatron reach at large $m_0$ is $\simeq 225$ GeV
in our BRpV model while it is of the order of 150 GeV in the MSSM.  Moreover,
unlike the trilepton signal, the discovery potential at small $m_0$ is also
increased with respect to that of the MSSM. In this region it is clear that
the reduction of the trilepton signal is largely due to the presence of
additional isolated leptons. As in Fig.\ \ref{fig:2}, the shaded area
represent the region where displaced vertices could be used to further
increase the sensitivity to BRpV.  In Fig.\ \ref{fig:3}(b) we present the
combined results for the trilepton and multilepton searches using the
chi-square criteria. It is
interesting to notice that the presence of $R$-parity violating interactions
leads to a $5\sigma$ SUSY discovery even at large $m_0$, a region where the
usual $R$-parity conserving SUGRA model has no discovery potential at all.
Notice the importance of combining trilepton and multilepton signals to
achieve this conclusion.

%%%% eps large tb small %%%%%%%%%%%%%%%

\begin{figure}
\begin{center}
\parbox[l]{3.5in}{
\mbox{\includegraphics[height=9cm,width=0.96\linewidth]{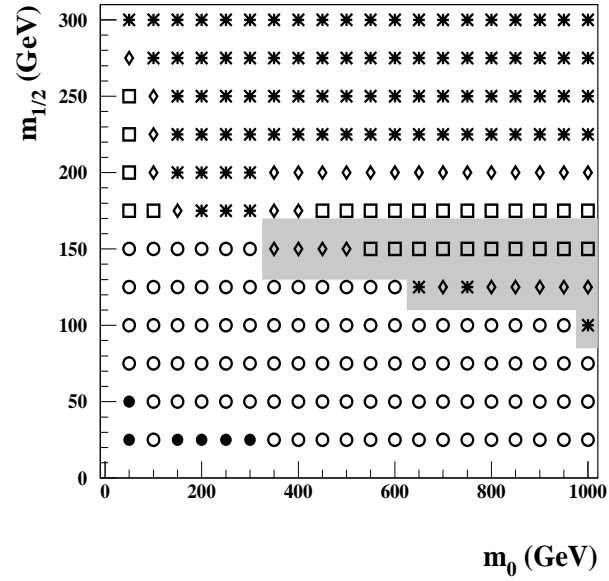}}
}
\hfill
\end{center}
\caption{
        Reach of Fermilab Tevatron Run II using the trilepton signal
        in the $m_0 \otimes m_{1/2}$ plane for $A_0=0$, $\tan\beta =
        3$, $\mu > 0$, $\epsilon_\tau = 0.22$ GeV, and $m_{\nu_3}
        = 0.05$ eV. The conventions are the ones in Fig.\ \ref{fig:2}.
}
\label{fig:6}
\end{figure}

\begin{figure}
\begin{center}
\parbox[l]{3.5in}{
\mbox{\includegraphics[height=9cm,width=0.96\linewidth]{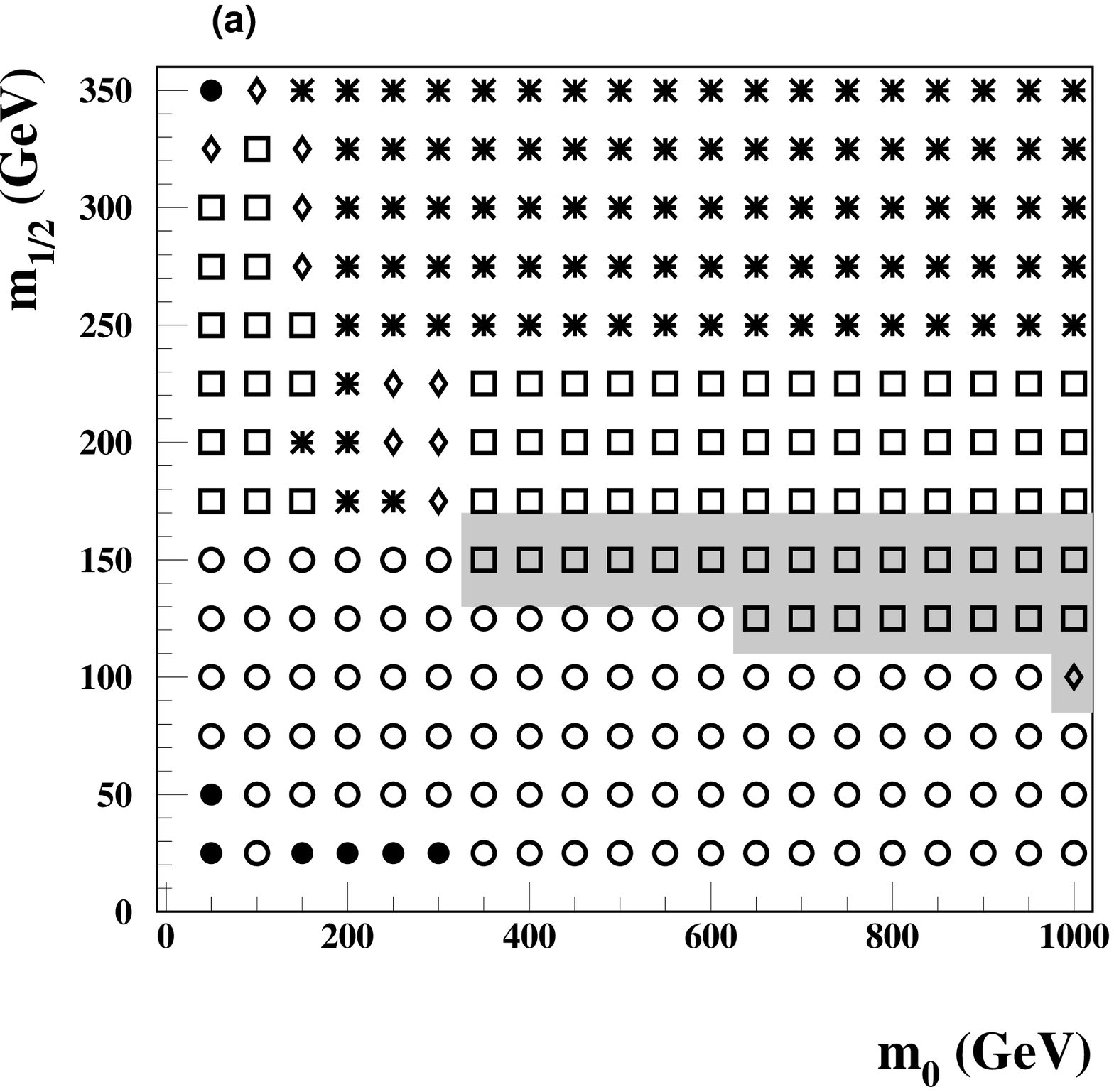}}
}
\hfill
\parbox[l]{3.5in}{
\mbox{\includegraphics[height=9cm,width=0.96\linewidth]{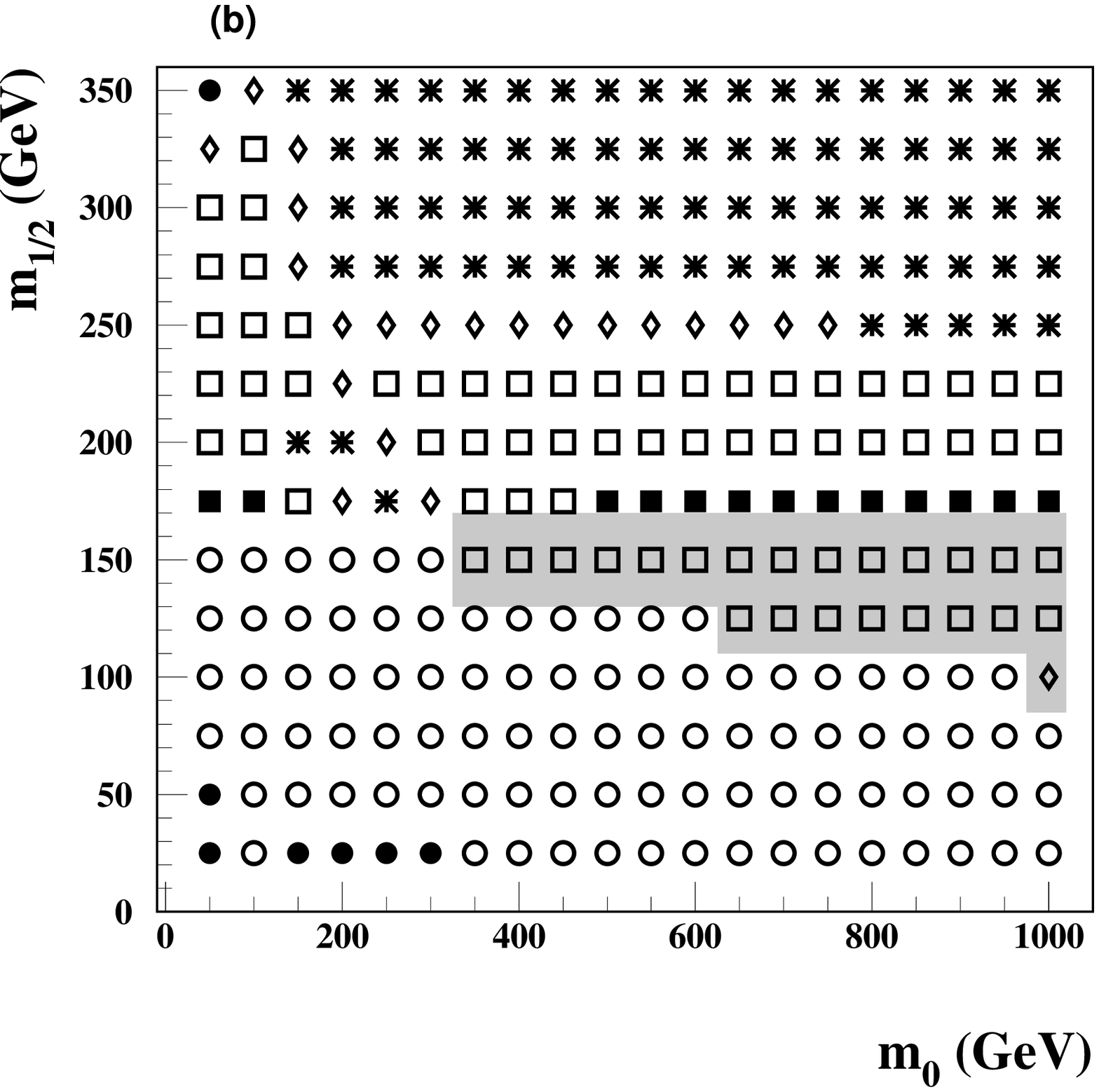}}
}
\end{center}
\caption{
        (a) Reach of Fermilab Tevatron Run II in the 4 or more lepton
        channel. (b) Combined trilepton and multilepton results. All
        parameters were chosen as in Fig.\ \ref{fig:6}.
}
\label{fig:8}
\end{figure}

Let us now consider a second scenario where the $\tilde{\chi}_1^0$ decays
predominantly into $\nu_3 b \bar{b}$ at low $m_0$. As seen in
Fig.\ \ref{fig:1.1}, the dominance of this decay channel happens for large
$\epsilon_\tau$ values since the LSP decay is scalar mediated.  The $\nu_3 b
\bar{b}$ decay mode spoils lepton isolation, without producing any additional
charged leptons and may be regarded, in a sense, as a worse case scenario in
comparison with the case of small $\epsilon_\tau$ and $m_0$.  In order to
illustrate this case we fixed $\epsilon_\tau = 0.22$ GeV and the remaining
parameters as before; $A_0=0$, $\tan \beta = 3$, $\mu > 0$, and $m_{\nu_3} =
0.05$ eV.  From Fig.\ \ref{fig:6} we can see that the trilepton reach is
indeed further reduced at small and medium $m_0$, as expected.  The same
happens for the multilepton signal; see Fig.\ \ref{fig:8}.  Nevertheless,
combining these signals still leads to a reach that is better than the MSSM
one for all $m_0$ and $m_{1/2}$ values.

%%%% eps small tb large %%%%%%%%%%%%%%%

\begin{figure*}[thbp] 
\includegraphics[height=9cm,width=0.46\linewidth]{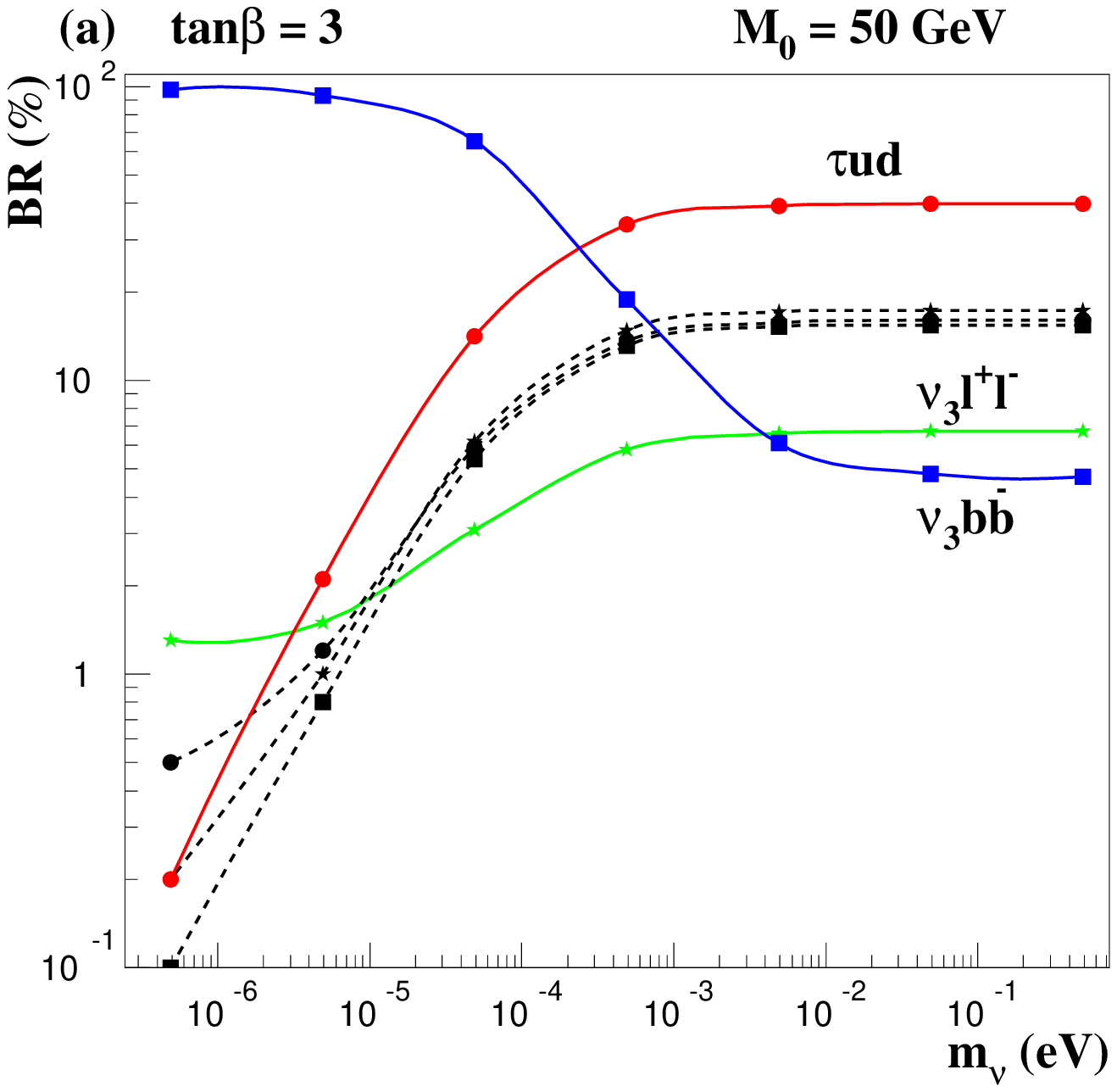}
\includegraphics[height=9cm,width=0.46\linewidth]{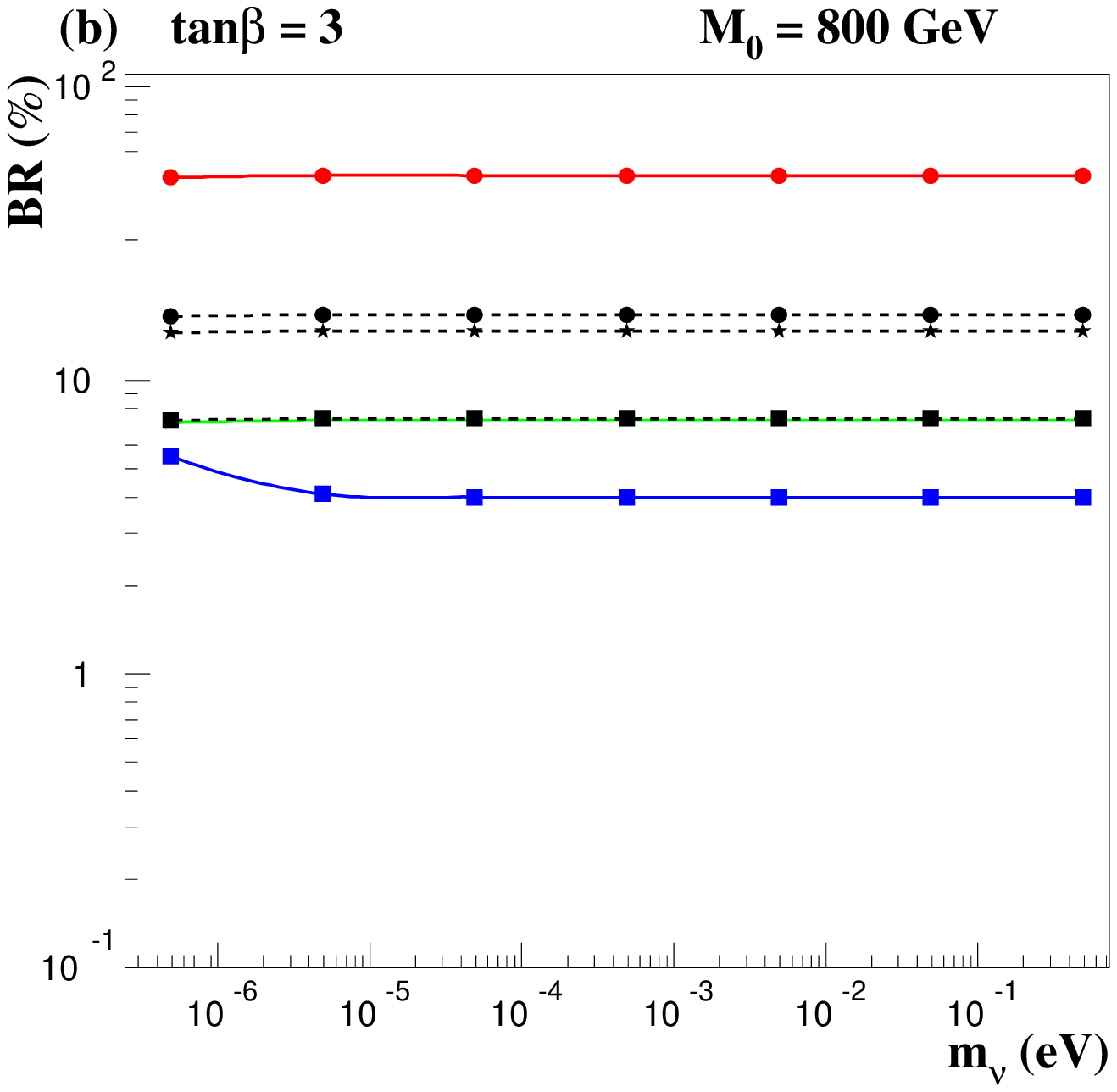}
\includegraphics[height=9cm,width=0.46\linewidth]{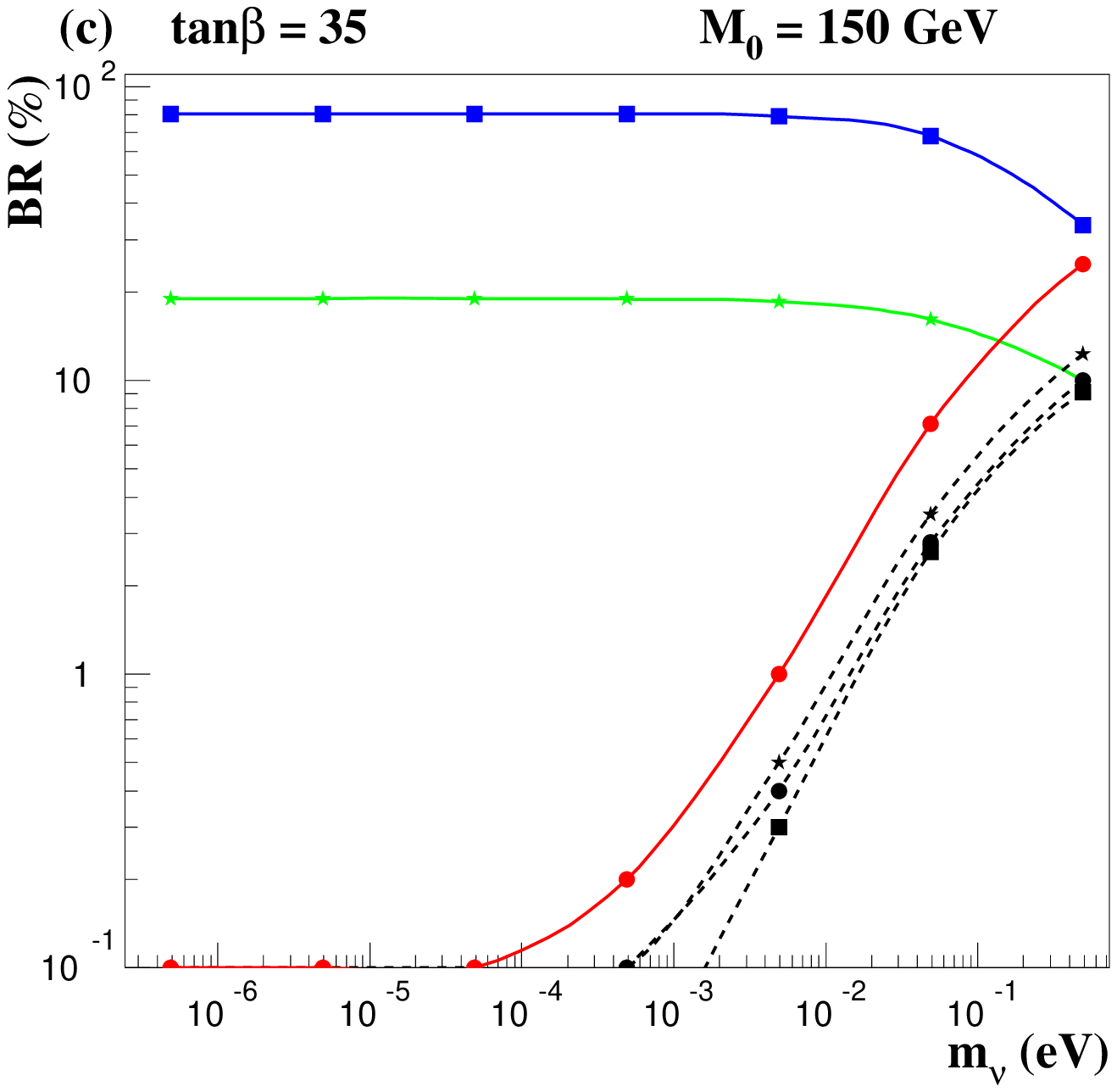}
\includegraphics[height=9cm,width=0.46\linewidth]{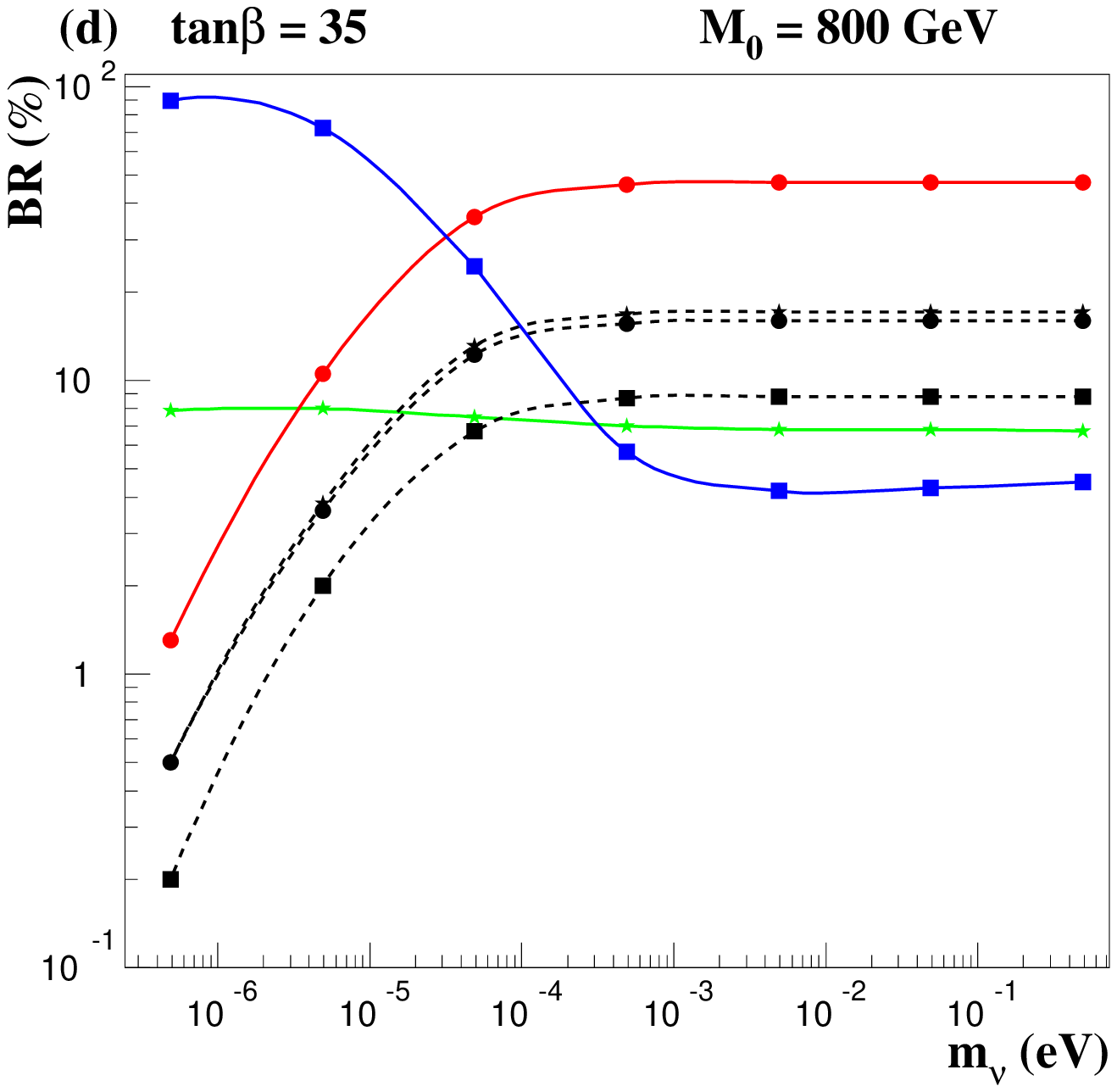}
\caption{
  $\tilde{\chi}_1^0$ branching ratios as a function of $m_{\nu_3}$ for
  $A_0 =0$, $\mu > 0$, and $\epsilon_\tau = 7\times 10^{-4}$ GeV. We
  fixed $m_{1/2} = 175$ GeV in the case of $\tan\beta = 3$ and
  $m_{1/2} = 125$ GeV when $\tan\beta = 35$. The lines are as in Fig.\
  \ref{fig:1}.  }
\label{fig:1.2}
\end{figure*}

\begin{figure}
\begin{center}
\parbox[l]{3.5in}{
\mbox{\includegraphics[height=9cm,width=0.96\linewidth]{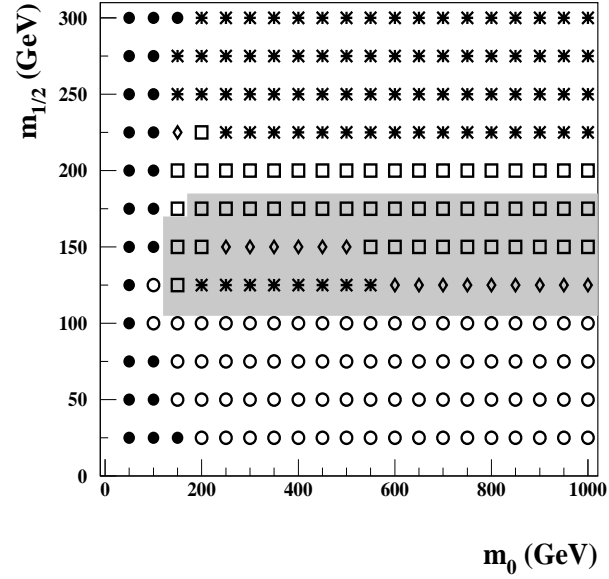}}
}
\end{center}
\caption{
        Reach of Fermilab Tevatron Run II using the trilepton signal
        in the $m_0 \otimes m_{1/2}$ plane for $A_0=0$, $\tan\beta =
        35$, $\mu > 0$, $\epsilon_\tau = 7\times 10^{-4}$ GeV, and $m_{\nu_3}
        = 0.05$ eV. The conventions are as in Fig.\ \ref{fig:2}.
}
\label{fig:4}
\end{figure}

Finally, we consider the case of small $\epsilon_\tau$ and large $\tan\beta$,
say $\epsilon_\tau = 7 \times 10^{-4}$ GeV and $\tan\beta = 35$.  Fig.\ 
\ref{fig:4} displays the Tevatron reach in the trilepton channel for this
case, keeping the remaining parameters as before; $A_0=0$, $\mu > 0$, and
$m_{\nu_3} = 0.05$ eV.  For these parameters, the main $\tilde{\chi}_1^0$
decay mode is $\tau u \bar d$.  However, there is a sizable contribution of
the $\nu b \bar{b}$ channel at small $m_0$.  As expected, the SUSY reach
decreases at small $m_0$, specially as we increase $\tan\beta$.  In contrast,
there is a slight gain for $m_0 \gtrsim 200$ GeV; see Fig.\ \ref{fig:4}.  The
similarity between the results in Fig.\ \ref{fig:2} and Fig.\ \ref{fig:4} at
large $m_0$ can be ascribed to the $\tau u d$ dominance of the LSP decay; see
Fig.\ \ref{fig:1.2}(b) and Fig.\ \ref{fig:1.2}(d).  In contrast the situation
is more complicated at smaller $m_0$ as seen in Fig.\ \ref{fig:2} and Fig.\ 
\ref{fig:4}.

\begin{figure}
\begin{center}
\parbox[l]{3.5in}{
\mbox{\includegraphics[height=9cm,width=0.96\linewidth]{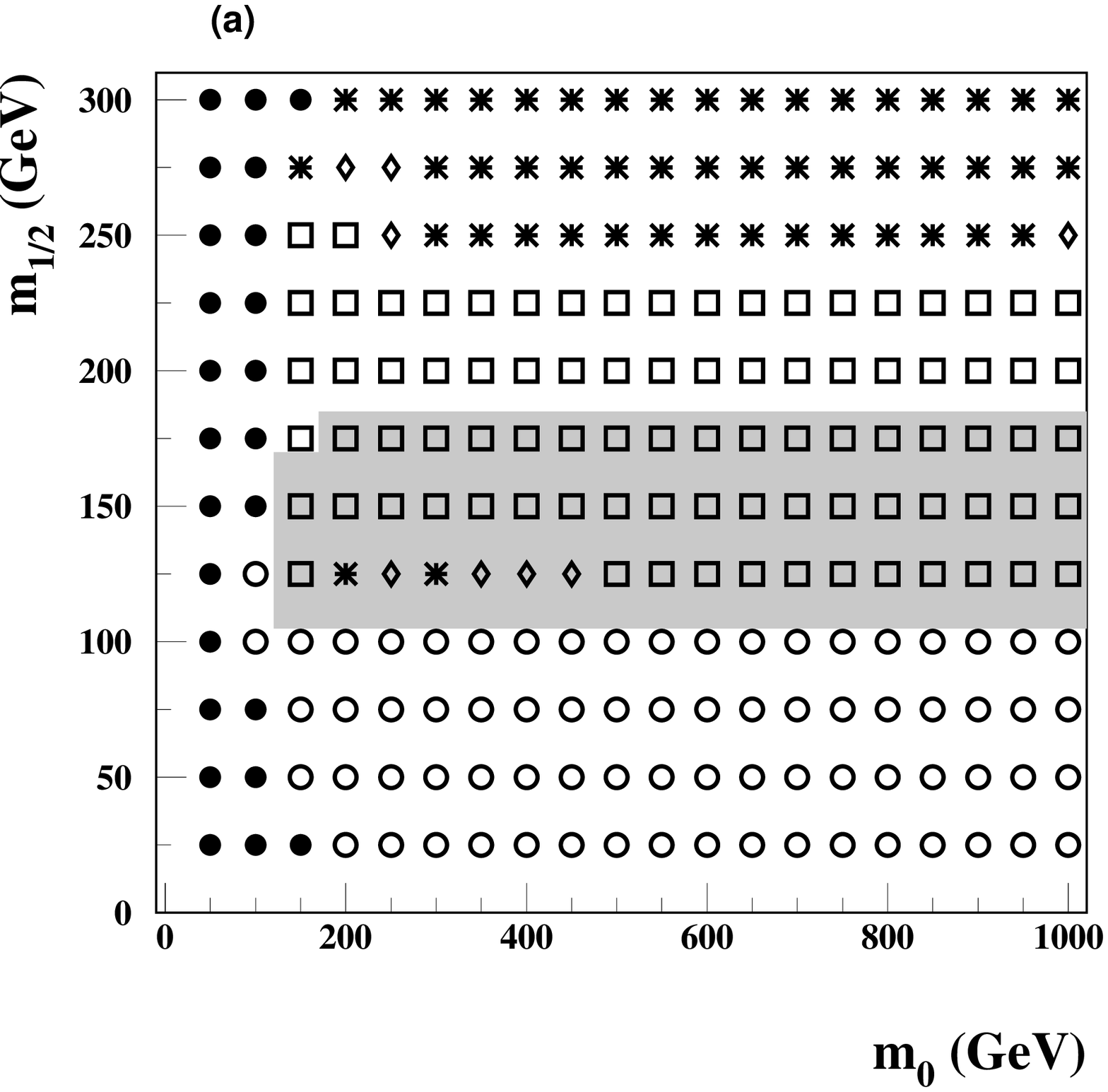}}
}
\hfill
\parbox[l]{3.5in}{
\mbox{\includegraphics[height=9cm,width=0.96\linewidth]{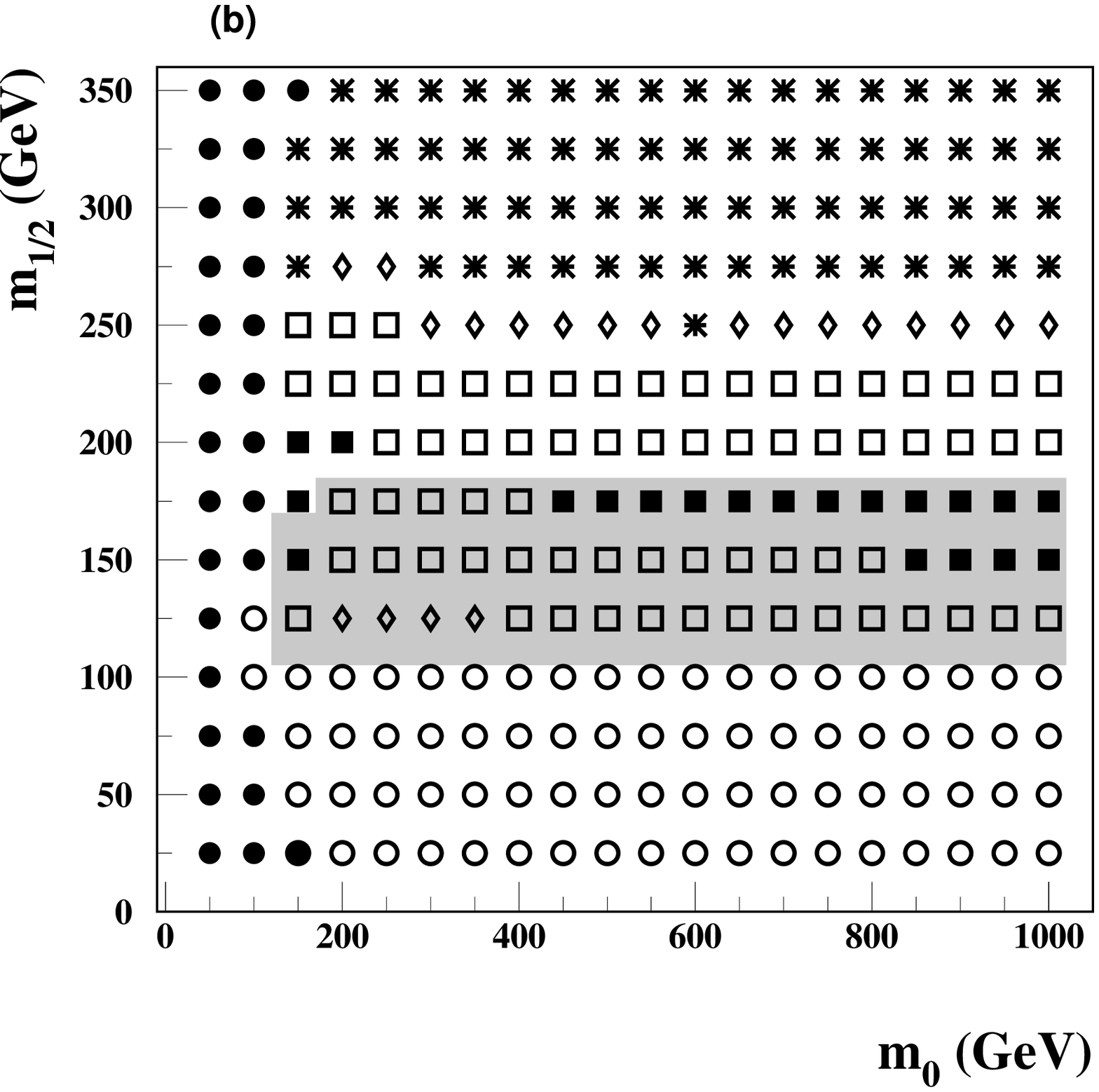}}
}
\end{center}
\caption{
        (a) Reach of Fermilab Tevatron Run II in the 4 or more lepton
        channel. (b) Combined trilepton and multilepton results. All
        parameters were chosen as in Fig.\ \ref{fig:4}.
}
\label{fig:5}
\end{figure}

For the last choice of parameters, the Tevatron discovery potential in the
multilepton channel is larger in our BRpV model than in the MSSM except at
small $m_0$, however, it is similar to the first case we analyzed; see Fig.\ 
\ref{fig:5}(a) and Fig.\ \ref{fig:3}(a).  This feature survives the
combination of three and multilepton signals, as can be seen from Fig.\ 
\ref{fig:3}(b) and Fig.\ \ref{fig:5}(b).  In contrast, the combined analysis
does make a difference in the small $m_0$ region.  In all cases, however, the
reach of the BRpV model is larger than the MSSM one.

%%%%%%%%%%%%%%%%%%%%%%%%%%%%%%%%%%%%%%%%%%%%%%%%%%%%%%%%%%%%%%%%%%%%%%
\section{Conclusion}

We have analyzed the production of multileptons ( $\ge 3 \ell$ with
$\ell = e$ or $\mu$) in the simplest supergravity model with violation
of $R$ parity at the Fermilab Tevatron. In this model, an effective
bilinear term in the superpotential parameterizes the explicit
breaking of $R$ parity.  Despite the small $R$-parity violating
couplings needed to generate the neutrino masses indicated by current
atmospheric neutrino data, the lightest supersymmetric particle is
unstable and can decay inside the detector. This leads to a
phenomenology quite distinct from that of the $R$-parity conserving
scenario.  We have quantified by how much the supersymmetric
multilepton signals differ from the $R$-parity conserving
expectations, displaying our results in the $m_0 \otimes m_{1/2}$
plane. We have shown that the presence of bilinear $R$-parity
violating interactions enhances the supersymmetric multilepton signals
over most of the parameter space, specially at moderate and large
$m_0$.  These topologies are useful not only for discovery, but also
to verify whether $R$ parity is conserved or not.

Adopting the hadronization procedures used in PYTHIA, we have first reproduced
the results for the trilepton signal expected in the conventional $R$-parity
conserving supergravity model.  We have found good agreement with the results
of Ref.~\cite{tata1} which adopts the ISAJET event generator. We have then
shown how the presence of BRpV interactions leads to a small suppression of
the trilepton signal at small values of $m_0$ irrespective of the value of
BRpV parameter $\epsilon_\tau$.  This is due to the $\tilde{\chi}_1^0$ decay
into $\nu b \bar{b}$.  However, the $\tilde{\chi}_1^0$ decays lead to a
drastically extended reach at large $m_0$ as a result of the LSP decay into
$\tau u \bar d$.  Moreover, the presence of additional isolated leptons in the
signal allows us to look for multilepton events.

We have demonstrated that combining the trilepton and multilepton
searches increases the Tevatron Run II sensitivity for most of SUGRA
and $R$-parity breaking parameters.  Note, however, that for neutrino
masses in the range indicated by current atmospheric data one has a
gap between the $\tilde{\chi}_1^0$ masses that can be probed at LEP2
(up to 40 GeV or so) and those that can be studied at the Tevatron
(above 70 GeV or so): within this range the $\tilde{\chi}_1^0$ decay
length is rather large, requiring the study of other topologies, like
the presence of displaced vertices in the tracking system.  It is
interesting to notice that we can search for SUSY signals also in the
low $m_0$ region by looking for events exhibiting multi jets + lepton
+ missing transverse momentum \cite{Barger:2001xe,Matchev:1999nb}.

In the present paper we have confined ourselves to the case in which
the lightest neutralino is also the lightest supersymmetric
particle~\footnote{In our model, any supersymmetric particle is a
  possible LSP candidate, since it is unstable and thus is not a
  cosmological relic.  However, only the lightest neutralino leads to
  a signal cross section which can be large enough to be measurable at
  the Tevatron.}, the most likely possibility if we adopt the simplest
set of supersymmetry soft breaking terms, universal at some
unification scale. We also have focused on the case where we have only
one generation and this is chosen to be the third. This is done first
for simplicity. Second we adopt this choice as a worse-case scenario.
In other words, in those parameter regions where our multi-lepton
signal can be discovered in the present one-generation approximation,
the inclusion of additional generations can only improve our result.
In contrast, in those regions where our results are negative, the
situation is totally inconclusive in the sense that a full fledged
analysis including all generations might reveal that the signal can
also be detected in part of those regions. Therefore our results are
robust, in the sense that the inclusion of additional generations
would imply new sources of leptons, specially muons.  The analysis is
substantially more involved, however, than the one considered here and
will be taken up elsewhere.

%%%%%%%%%%%%%%%%%%%%%%%%%%%%%%%%%%%%%%%%%%%%%%%%%%%%%%%%%%%%%%%%%%%%%%

\begin{acknowledgments}
  
  Research supported by Funda\c{c}\~{a}o de Amparo \`a Pesquisa do
  Estado de S\~ao Paulo (FAPESP), by Conselho Nacional de
  Desenvolvimento Cient\'{\i}fico e Tecnol\'ogico (CNPq), by Programa
  de Apoio a N\'ucleos de Excel\^encia (PRONEX), by Spanish MCyT grant
  BFM2002-00345 and by the European Commission RTN network
  HPRN-CT-2000-00148. W.P.~has been supported by the 'Erwin
  Schr\"odinger fellowship' No. J2272
  of the 'Fonds zur F\"orderung der wissenschaftlichen Forschung' of Austria
  and partly by the Swiss 'Nationalfonds'. D.R.~has been partially
  supported by UdeA-CODI Grant No. 381-8-10-02.
  
\end{acknowledgments}

%%%%%%%%%%%%%%%%%%%%%%%%%%%%%%%%%%%%%%%%%%%%%%%%%%%%%%%%%%%%%%%%%%%%%%

%%%%%%%%%%%%%%%%%%%%%%%%%%%%%%%%%%%%%%%%%%%%%%%%%%%%%%%%%%%%%%%%%%%%%%
\newpage
\begin{table}
\begin{center}
\begin{tabular}{|l|c|}\hline
BG (fb)  & $\sigma$ (fb)  \\ \hline \hline
WZ (Z $\to  \tau \tau$)  & 0.17  \\ \hline
W$^*$Z$^*$,W$^*$$\gamma\to  ll\bar{l}$  & 0.12  \\ \hline
W$^*$Z$^*$,W$^*$$\gamma\to  ll^{'}\bar{l^{'}}$  & 0.15  \\ \hline
$t \bar{t}$  & 1.15  \\ \hline
Z$^*$Z$^*$  & 0.05  \\\hline
total  & 1.64  \\
\hline \hline
\end{tabular}
\end{center}
\caption{Background cross sections in fb for the trilepton signal at the
Tevatron Run II after kinematical cuts discussed in the text.}
\label{table1}
\vskip 0.3cm
\end{table}

%%%%

\begin{table}
\begin{center}
\begin{tabular}{|l|c|}\hline
BG (fb)  & $\sigma$ (fb) \\ \hline \hline
WZ   & 0.01  \\ \hline
Z$^*$Z$^*$  & 0.10  \\\hline
$t \bar{t}$  & 0.16  \\ \hline
total  & 0.27   \\
\hline \hline
\end{tabular}
\end{center}
\caption{Background cross sections in fb for the multilepton signal at the
Fermilab Tevatron Run II after kinematical cuts discussed in the text.}
\label{table2}
\end{table}

%\newpage
%%%%%%%%%%%%%%%%%%%%%%%%%%%%%%%%%%%%%%%%%%%%%%%%%%%%%%%%%%%%%%%%%%%%%%

%%%

%%%%

% \begin{figure}
% \begin{center}
%   \includegraphics[height=15cm,width=0.96\linewidth]{br_eps.eps}
% \end{center}

 %%%%

% \begin{figure}
% \begin{center}
% \includegraphics[height=15cm,width=0.96\linewidth]{br_mnu.eps}
% \end{center}

%%%%

%%%%

%%%%%

%%%

%%%%

%%%%

%%%%%%%%%%%%%%%%%%%%%%%%%%%%%%%%%%%%%%%%%%%%%%%%%%%%%%%%%%%%%%%%%%%%%%


\begin{thebibliography}{99}
\bibitem{3l-4l} In the trilepton analysis, we reject events presenting
more than three isolated charged leptons.  In the $R$-parity conserving case,
this procedure makes no difference.
\bibitem{tatadif} This procedure differs from the one used in Ref.\
\cite{tata1}, which used only Gaussian statistics. The use of Poison
statistics leads to more restrictive bounds.
\end{thebibliography}
\end{document}